\documentclass[10pt]{article}

\usepackage[utf8]{inputenc}
\usepackage[T1]{fontenc}
\usepackage{arxiv}
\usepackage{xr}
\usepackage{url}
\usepackage{booktabs}
\usepackage{amsmath, amssymb, amsfonts, amsthm}
\usepackage{nicefrac}
\usepackage{microtype}
\usepackage{graphicx}
\usepackage{xcolor}
\usepackage{hyperref}
\usepackage[nameinlink,capitalize]{cleveref}
\usepackage[numbers,sort&compress]{natbib}
\usepackage{algorithm}
\usepackage{algpseudocode}
\usepackage{caption}
\usepackage{subcaption}
\usepackage{siunitx}
\usepackage{xspace}

\crefname{figure}{Fig.}{Figs.}
\Crefname{figure}{Fig.}{Figs.}
\crefname{section}{Sec.}{Secs.}
\Crefname{section}{Sec.}{Secs.}
\crefname{algorithm}{Algorithm}{Algorithms}
\Crefname{algorithm}{Algorithm}{Algorithms}
\crefname{table}{Table}{Tables}
\Crefname{table}{Table}{Tables}
\crefname{equation}{equation}{equations}
\Crefname{equation}{Equation}{Equations}

\newcommand{\ie}{i.e.,\xspace}
\newcommand{\eg}{e.g.,\xspace}
\newcommand{\Norm}[1]{\left\lVert#1\right\rVert}

\newcommand{\etalcite}[1]{et al.~\cite{#1}}

\providecommand{\articletype}[1]{}

\providecommand{\orcid}[1]{}

\title{Neural Spectral Element Methods for stiff multiphysics PDEs
       with electrochemical transport benchmarks}

\author{%
  \begin{minipage}[t]{0.48\linewidth}
    \centering
    \textbf{Conrard Giresse Tetsassi Feugmo}$^{\,1,2,*}$\\[2pt]
    {\small $^{1}$Department of Chemistry\\
            $^{2}$Department of Physics and Astronomy\\
            University of Waterloo\\
            200 University Ave.\ West\\
            Waterloo, ON N2L 3G1, Canada}\\[2pt]
    {\small$^{*}$\texttt{cgtetsas@uwaterloo.ca}}
  \end{minipage}\hfill
  \begin{minipage}[t]{0.48\linewidth}
    \centering
    \textbf{David Pankaczy}$^{\,2}$\\[2pt]
    {\small $^{2}$Department of Physics and Astronomy\\
            University of Waterloo\\
            200 University Ave.\ West\\
            Waterloo, ON N2L 3G1, Canada}
  \end{minipage}%
}

\date{}

\providecommand{\ack}[1]{\section*{Acknowledgements}#1}
\providecommand{\funding}[1]{\section*{Funding}#1}
\providecommand{\roles}[1]{\section*{Author contributions}#1}
\providecommand{\data}[1]{\section*{Data availability}#1}
\providecommand{\suppdata}[1]{\section*{Supplementary information}#1}

\begin{document}

\maketitle

\begin{abstract}
\noindent
Physics-informed neural networks (PINNs) are limited by three
interlocking bottlenecks, the \emph{PINN trilemma}: Monte-Carlo
collocation, serial automatic differentiation, and a resulting
stochastic loss landscape that prevents quasi-Newton optimisation.
These bottlenecks place an empirical $\mathcal{O}(10^{-2})$ accuracy
floor on stiff multiphysics problems such as electrochemical
transport, well above the sub-percent precision required for
quantitative parameter inference. In this paper we address that
accuracy floor by replacing the random-collocation pipeline with a
spectral one. We introduce the Neural Spectral Element Method
(NSEM), which evaluates each network only at fixed
Legendre--Gauss--Lobatto quadrature nodes and replaces all derivative
calls with precomputed spectral differentiation matrices. The
resulting deterministic loss enables limited-memory BFGS (L-BFGS) to reach residuals of
$10^{-9}$--$10^{-10}$. A Kosloff--Tal-Ezer coordinate map resolves
electrochemical boundary layers; a mesh-free neural mortar framework
couples multi-element domains. On the four-example Poisson--Nernst--Planck (PNP) benchmark of
Huang and co-workers, NSEM attains $10^{-4}$--$10^{-7}$ relative
pointwise error with two orders of magnitude fewer collocation points
than the adaptive-resampling PINN baseline. Both a tanh multilayer perceptron (MLP) and a
basis-aligned Legendre Kolmogorov--Arnold Network (KAN) backbone attain spectral accuracy within
the same NSEM infrastructure, with the KAN requiring roughly half the
Adam steps to enter the L-BFGS basin of attraction on the 1D PNP
benchmark.
\end{abstract}

\vspace{0.4em}
\noindent\textbf{Keywords:} neural spectral element methods;
  physics-informed neural networks; Poisson--Nernst--Planck;
  spectral collocation; Legendre--Gauss--Lobatto quadrature;
  mortar elements; Kolmogorov--Arnold networks; L-BFGS.

\vspace{0.8em}

\section{Introduction}\label{sec:intro}

Physics-informed neural networks (PINNs) embed the residual of a
partial differential equation as a soft penalty in the training loss,
requiring neither labelled data nor a computational mesh
\citep{raissi2019pinn,karniadakis2021piml}. Their flexibility has
spurred applications across fluid mechanics, materials science, and
electrochemistry, but three interlocking limitations --- which we call
the \emph{PINN trilemma} --- bound the accuracy that is practically
achievable. First, PDE residuals are integrated by
Monte-Carlo sampling over randomly re-drawn collocation points, which
converges at the slow rate $\mathcal{O}(N^{-1/2})$ and injects
high-frequency noise into the gradient even for smooth integrands,
contributing to the stiff-PDE failure modes documented by
Krishnapriyan~\etalcite{krishnapriyan2021failure}. Second, spatial derivatives are
evaluated by automatic differentiation through the network, at cost
$\mathcal{O}(L)$ in evaluation time for a depth-$L$ network and
$\mathcal{O}(L^{2})$ for a full Laplacian, while serialising the
computation graph in a way that underutilises batched linear-algebra
hardware \citep{tancik2020fourier,rahaman2019spectralbias}. Third, the
stochastic gradient produced by random resampling corrupts the curvature
information required by quasi-Newton optimisers such as the limited-memory BFGS method (L-BFGS), so
practitioners are restricted to first-order methods that plateau several
orders of magnitude above machine precision
\citep{mishra2023pinn,mishra2022inverse}.

We introduce the \emph{Neural Spectral Element Method} (NSEM), which
attacks all three legs of the trilemma simultaneously. The core substitution is
to evaluate the network only at the $N$ fixed LGL quadrature nodes and
to replace all autodiff derivative calls by left-multiplication with
the precomputed spectral matrices $\mathbf{D}^{(1)}$ and
$\mathbf{D}^{(2)}=\mathbf{D}^{(1)}\mathbf{D}^{(1)}$, assembled once
before training at $\mathcal{O}(N^{2})$ cost independent of network
depth. The physics loss thereby becomes a deterministic function of
the trainable parameters, enabling the four-phase training schedule
of \Cref{sec:method:training} to reach residuals below
$10^{-10}$. Multi-element domains are handled by a mesh-free neural mortar
framework (\Cref{sec:method:mortar}), and a Kosloff--Tal-Ezer
coordinate map resolves thin electrochemical boundary layers with no
extra node cost. We validate NSEM on a full four-example
Poisson--Nernst--Planck (PNP) benchmark suite,
spanning 1D and 3D steady problems and 1D and 2D time-dependent
problems with manufactured analytic solutions and stiff coupling
coefficients, and on three canonical convergence tests (Helmholtz,
Allen--Cahn, convection--diffusion). We compare two interchangeable
network backbones --- a tanh-activated MLP and a Legendre-KAN
\citep{liu2024kan} whose polynomial edge-functions are aligned with the
LGL basis (\Cref{sec:backbone}) --- and find that both achieve
spectral accuracy with identical solver infrastructure.

\section{Related work}\label{sec:related}

\paragraph{Classical spectral elements.}
The classical spectral element method (SEM) achieves exponential
convergence by representing each field on a polynomial basis supported
at Gauss--Lobatto quadrature nodes and coupling elements through mortar
projections \citep{patera1984spectral,karniadakis2005spectral}, and is
the method of choice for smooth PDEs in computational fluid dynamics
and seismology \citep{komatitsch1999spectral}.

\paragraph{Domain decomposition for PINNs.}
Several lines of work attack the multi-scale failure mode of PINNs
through geometric decomposition. XPINN \citep{jagtap2020xpinn}
assigns one network per non-overlapping subdomain and enforces value
and flux continuity through interface losses. FBPINN
\citep{moseley2023fbpinn} uses overlapping subdomains coupled by a
partition of unity. Variational PINNs and their hp-refined
descendants
\citep{kharazmi2021hpvpinn,anandh2024fastvpinn}
recast the residual loss in weak form with piecewise polynomial test
functions. These approaches reduce per-subdomain variance, but they
retain Monte-Carlo sampling for the residual integral and automatic
differentiation for derivatives; the trilemma still bites every
subdomain.

\paragraph{Spectral-neural couplings.}
Three concurrent works pursue orthogonal-polynomial differentiation
more aggressively. Du~\etalcite{du2024nsm} minimise a Parseval-norm loss in
a spectral expansion but do not use physical-space collocation or
multi-element geometry. Yu~\etalcite{yu2024sinn} replace automatic
differentiation by spectral multiplication in Fourier space but omit
boundary-layer adaptation. Shukla~\etalcite{shukla2024neurosem} couple a PINN
correction to a classical Nektar++ SEM solver, keeping an external
mesh and solver in the loop.

\paragraph{Residual reweighting and causal training.}
A second family of improvements modifies the loss aggregator rather
than the geometry. PIRBN \citep{anagnostopoulos2024pirbn} introduces
a residual-based attention weighting; causal training
\citep{wang2024causal} imposes an explicit temporal weighting that
respects the causal structure of time-dependent PDEs. Both methods
close part of the accuracy gap but do not remove the Monte-Carlo
quadrature or the autodiff derivative cost, so the L-BFGS-grade
landscape is still out of reach.

\paragraph{Kolmogorov--Arnold networks.}
KANs \citep{liu2024kan} replace fixed activations by learnable
univariate edge functions, originally parameterised as B-splines.
Polynomial-edge variants
\citep{wang2024kinn,zhang2024legendkinn,ss2024chebkan,sankaran2024chebpikan}
explore Legendre and Chebyshev bases; only the Legendre choice
combined with LGL quadrature delivers the exact basis-alignment
property we exploit in \Cref{sec:backbone:why_legendre}.

\paragraph{Positioning of NSEM.}
No prior method combines all five of the ingredients NSEM requires:
(i) deterministic physical-space quadrature at Legendre--Gauss--Lobatto
(LGL) nodes; (ii) precomputed spectral differentiation matrices;
(iii) nonlinear coordinate stretching for boundary-layer resolution;
(iv) a mesh-free multi-element mortar framework; and (v) L-BFGS polish
to machine precision on a stiff coupled PDE system. The remainder of
this paper develops these five ingredients and demonstrates that their
combination eliminates the PINN trilemma on stiff multiphysics
benchmarks.

\section{The Neural Spectral Element Method}\label{sec:method}

\begin{figure}[t]
 \centering
 \includegraphics[width=0.97\textwidth]{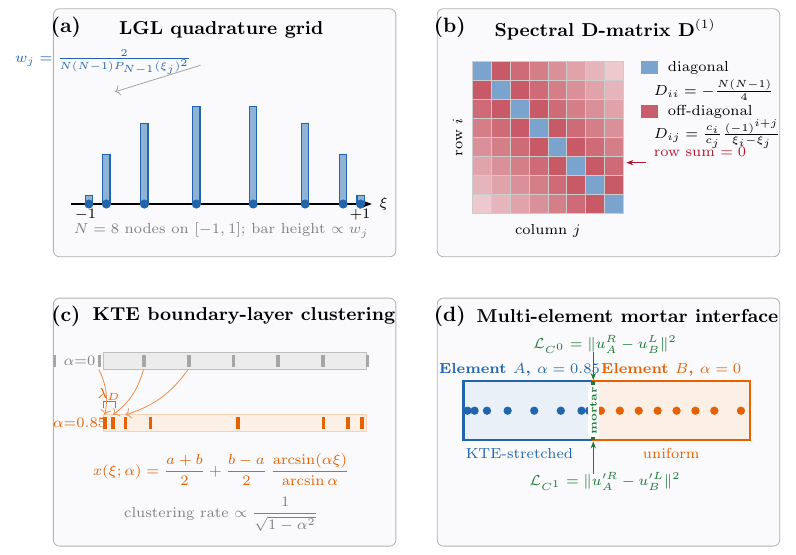}
 \caption{Architecture of the NSEM framework.
 \textbf{(a)} The $N$ Legendre--Gauss--Lobatto nodes on the reference
 element $[-1,1]$ with their quadrature weights $w_{j}$ drawn as bar
 heights; the network is evaluated only at these fixed nodes, making
 the loss in \cref{eq:nsem_loss} deterministic.
 \textbf{(b)} The spectral first-derivative matrix
 $\mathbf{D}^{(1)}$ assembled in barycentric form
 (\citealp{berrut2004barycentric}; \cref{eq:Dmat});
 $\mathbf{D}^{(2)}=\mathbf{D}^{(1)}\mathbf{D}^{(1)}$ replaces a
 full autodiff Laplacian by one dense general matrix--matrix multiplication (GEMM) per forward pass.
 \textbf{(c)} The Kosloff--Tal-Ezer map
 (\cref{eq:kte}) deforms the uniform LGL distribution
 towards either boundary, controlled by a single learnable parameter
 $\alpha$; the larger $\alpha$, the sharper the boundary-layer
 resolution.
 \textbf{(d)} Two non-conforming elements joined by an $L^{2}$ Lagrange
 mortar projection. Value continuity is exact by construction;
 the diffusive flux is enforced by a $C^{1}$ penalty
 (\Cref{sec:method:mortar}).}
 \label{fig:schematic}
\end{figure}

\subsection{LGL collocation grid}\label{sec:method:lgl}
\Cref{fig:schematic}(a) illustrates the LGL nodal grid that underpins
this section.
A standard PINN approximates the spatial integral
$\int_{\Omega}|\mathcal{R}(x)|^{2}\,\mathrm{d}x$ by a Monte-Carlo
average over randomly re-drawn samples, which converges at the slow
rate $\mathcal{O}(N^{-1/2})$ and introduces stochastic gradient noise
that prevents convergence of second-order optimisers
\citep{raissi2019pinn,karniadakis2021piml}. Replacing this estimator
by a Legendre--Gauss--Lobatto (LGL) quadrature rule yields a
deterministic loss with exponential convergence for analytic
integrands, the standard accuracy guarantee of spectral element
methods \citep{patera1984spectral,karniadakis2005spectral,trefethen2000spectral,boyd2001spectral}.
On the reference interval $[-1,1]$ the LGL nodes consist of the two
endpoints $\xi_{0}=-1$ and $\xi_{N-1}=+1$ together with the $N-2$
interior zeros of $P_{N-1}'$, the derivative of the
$(N-1)$-th Legendre polynomial; we compute these zeros by
Newton--Raphson iteration seeded from Chebyshev nodes. The
associated quadrature weights are
$w_{j}=2\bigl/\bigl[N(N-1)\,P_{N-1}(\xi_{j})^{2}\bigr]$
and define the discrete physics loss
\begin{equation}
 \mathcal{L}_{\mathrm{NSEM}}
 \;=\;\sum_{j=0}^{N-1} w_{j}\,\bigl|\mathcal{R}(\xi_{j})\bigr|^{2}
 \;\approx\;\Norm{\mathcal{R}}_{L^{2}(\Omega)}^{2}.
 \label{eq:nsem_loss}
\end{equation}
Because the nodes $\{\xi_{j}\}$ are fixed for the entire training
run, the gradient $\nabla_{\theta}\mathcal{L}_{\mathrm{NSEM}}$ has
exactly zero estimator variance. The $N$-point LGL rule integrates
any polynomial of degree $\leq 2N-3$ exactly, and the quadrature
error decays as $|E_{N}|\leq C\,e^{-\sigma N}$ for analytic
integrands \citep{karniadakis2005spectral,boyd2001spectral}; at
$N=32$ the residual integration error is already at machine
precision.

\subsection{Spectral differentiation matrix}\label{sec:method:dmatrix}
The structure of the precomputed $\mathbf{D}^{(1)}$ matrix is sketched
in \Cref{fig:schematic}(b).
A standard PINN obtains derivatives by backpropagating through the
network, which costs $\mathcal{O}(L)$ per evaluation for a network of
depth $L$ for a first derivative and $\mathcal{O}(L^{2})$ for a
Laplacian, and serialises the computation graph in a way that
underutilises modern GPU tensor cores. The NSEM forward pass replaces
both autodiff calls by a single dense matrix--vector product.
Evaluating the network only at the LGL nodes yields a vector
$\mathbf{u}=[\hat u(\xi_{0}),\dots,\hat u(\xi_{N-1})]^{\top}$; the
spectral first-derivative matrix $\mathbf{D}^{(1)}$ then provides
the discrete Lagrange derivative at every node by
$\mathbf{u}^{(1)}=\mathbf{D}^{(1)}\mathbf{u}$. We assemble
$\mathbf{D}^{(1)}$ in the barycentric form of
Berrut and Trefethen~\cite{berrut2004barycentric},
\begin{equation}
 D^{(1)}_{ij}
 \;=\;\frac{c_{i}}{c_{j}}\cdot\frac{(-1)^{i+j}}{\xi_{i}-\xi_{j}}
 \quad(i\neq j),
 \qquad
 D^{(1)}_{00}=-\frac{N(N-1)}{4},
 \;\; D^{(1)}_{N-1,N-1}=+\frac{N(N-1)}{4},
 \;\; D^{(1)}_{ii}=0\;\,(0<i<N-1),
 \label{eq:Dmat}
\end{equation}
with endpoint corrections $c_{0}=c_{N-1}=2$ and $c_{i}=1$ otherwise.
The second-derivative operator is then $\mathbf{D}^{(2)}=\mathbf{D}^{(1)}\mathbf{D}^{(1)}$,
computed once before training and stored as a constant tensor. The
construction is the polynomial-collocation analogue of the discrete
variable representation in chemical physics
\citep{light1985dvr,colbert1992dvr}: it is exact on the polynomial
subspace of degree $\leq N-1$, and converges exponentially in $N$ on
analytic functions. This is the appropriate response to the recent
defence of autodiff for PDE solving by
Wang~\etalcite{wang2024adessential}: their argument assumes finite-difference
or random-sample discretisations whose own errors dominate; in NSEM
the LGL inner product is polynomial-exact up to degree $2N-3$, so the
spectral matrix is exact on the relevant subspace and adding autodiff
on top would merely re-derive the same numbers at higher cost. A
practical caveat is essential when the LGL grid is composed with a
nonlinear coordinate map: the physical second-derivative matrix is
$\mathbf{D}^{(2)}_{\mathrm{phys}}=\mathbf{D}^{(1)}_{\mathrm{phys}}\mathbf{D}^{(1)}_{\mathrm{phys}}$
and \emph{not} $\mathrm{diag}(1/J^{2})\,\mathbf{D}^{(2)}_{\mathrm{ref}}$,
because the Jacobian $J(\xi)$ is itself $\xi$-dependent (see
\Cref{sec:method:kte}); applying the naive formula degrades
the steady nonlinear Gouy--Chapman benchmark to a relative error of
$\sim\!6\%$, whereas the correct chain-rule product yields $\sim\!0.15\%$.

\subsection{Kosloff--Tal-Ezer coordinate mapping}\label{sec:method:kte}
The KTE coordinate stretching used in this subsection is depicted in
\Cref{fig:schematic}(c).
LGL nodes cluster near each element boundary at rate $\mathcal{O}(N^{-2})$,
which is insufficient for problems whose physical solution varies on a
much shorter length scale --- the Debye layer in electrochemistry, for
instance, has thickness $\lambda_{D}\sim\kappa^{-1}\ll N^{-2}$ in the
non-dimensional cell. We resolve such boundary layers by composing
the LGL grid with the analytic stretching of
Kosloff and Tal-Ezer~\cite{kosloff1993kte},
\begin{equation}
 x(\xi;\alpha)
 \;=\;\frac{a+b}{2}\;+\;\frac{b-a}{2}\,
 \frac{\arcsin(\alpha\xi)}{\arcsin(\alpha)},
 \qquad \alpha\in[0,1),
 \label{eq:kte}
\end{equation}
which sends $\xi\in[-1,1]$ to $x\in[a,b]$ and reduces to the linear
affine map as $\alpha\to 0$ while concentrating nodes exponentially
near $x=a$ and $x=b$ as $\alpha\to 1$. The Jacobian is
$J(\xi)=\bigl[(b-a)/(2\arcsin\alpha)\bigr]\alpha/\sqrt{1-\alpha^{2}\xi^{2}}$,
and the physical first-derivative matrix is the row-scaling
$\mathbf{D}^{(1)}_{\mathrm{phys}}=\mathrm{diag}(1/J)\,\mathbf{D}^{(1)}_{\mathrm{ref}}$;
the physical quadrature weights become
$w_{j}^{\mathrm{phys}}=J(\xi_{j})\,w_{j}^{\mathrm{ref}}$. All KTE
rescalings are precomputed once and stored, so the stretching adds no
per-step cost. We expose $\alpha$ as a learnable parameter when the
optimal clustering is not known a priori, which is the usual case for
non-self-similar boundary layers. \Cref{fig:schematic}(c)
illustrates the effect on node placement.

\subsection{Multi-element mortar interfaces}\label{sec:method:mortar}
The two-element mortar configuration with $C^{0}$ and $C^{1}$
interface penalties is sketched in \Cref{fig:schematic}(d).
A single KTE-mapped element resolves one boundary layer but not two
disparate scales separated by a bulk region; for those cases we
decompose the physical domain into $K$ disjoint subdomains
$\Omega^{(e)}$ and assign each a private network whose outputs are
LGL-collocated as in \Cref{sec:method:lgl}. Following the
classical mortar method \citep{bernardi1994mortar,lacour1997mortar},
continuity between non-conforming neighbouring elements is enforced
weakly by $L^{2}$ Lagrange projection between LGL traces. Because
the mass matrix associated with LGL nodes is diagonal, the projection
matrix between two-element traces reduces to a single sparse
matrix--vector product per training step, with no additional storage
or solve. We supplement this $C^{0}$ value continuity with an
explicit $C^{1}$ flux-continuity penalty that compares the spectral
derivative on the two sides of each interface, yielding a soft
constraint on the diffusive flux which is essential for stiff
transport problems with a thin space-charge layer at the joint. This
non-conforming, disjoint-element design contrasts with the
overlapping-window construction of FBPINNs \citep{moseley2023fbpinn}
and the residual-continuity approach of XPINNs
\citep{jagtap2020xpinn}: NSEM elements share no points, no
partition-of-unity weights are needed, and global continuity is
enforced by the mortar projection rather than by a smoothing window.

\subsection{Loss assembly}\label{sec:method:loss}
The total training objective combines the deterministic PDE residual
with boundary and interface penalties:
\begin{equation}
 \mathcal{L}\;=\;\sum_{e=1}^{K}\bigl(\widetilde{w}^{(e)}\bigr)^{\!\top}\!\bigl(\mathbf{R}^{(e)}\odot\mathbf{R}^{(e)}\bigr)
 \;+\;\lambda_{\mathrm{BC}}\,\mathcal{L}_{\mathrm{BC}}
 \;+\;\lambda_{\mathrm{C}^{0}}\,\mathcal{L}_{\mathrm{C}^{0}}
 \;+\;\lambda_{\mathrm{C}^{1}}\,\mathcal{L}_{\mathrm{C}^{1}},
 \label{eq:total_loss}
\end{equation}
where $\mathbf{R}^{(e)}$ is the residual vector evaluated at the LGL
nodes of element $e$, $\mathcal{L}_{\mathrm{BC}}$ collects Dirichlet,
Neumann, and Robin (\eg Butler--Volmer) boundary contributions, and
$\mathcal{L}_{\mathrm{C}^{0}}, \mathcal{L}_{\mathrm{C}^{1}}$ enforce value and flux
continuity at element interfaces (\Cref{sec:method:mortar}).
The per-element normalisation
$\widetilde{w}^{(e)}=w^{(e)}\bigl/\mathbf{1}^{\!\top}w^{(e)}$ is
mandatory when subdomains of very different physical size coexist;
without it the raw LGL weights scale with the element length and the
PDE loss is dominated by the largest element, so the network never
resolves the thin layer. We observed this failure mode for the
charged-wall problem with a domain-size ratio of $5000{:}1$ between
bulk and EDL elements --- with raw weights, the bulk contribution
exceeds the interface contribution by three orders of magnitude and
L-BFGS stalls after one to two steps; the per-element normalisation
restores convergence. The constant multipliers
$(\lambda_{\mathrm{BC}},\lambda_{\mathrm{C}^{0}},\lambda_{\mathrm{C}^{1}})$ can be set
manually, but for problems with heterogeneous loss scales we instead
employ the balanced residual-decay-rate (BRDR) adaptive aggregator
\citep{chen2024brdr} (Supplementary~Sec.~S1.5), which re-weights each
loss component to equalise relative decay rates via an exponential
moving average ($\beta=0.999$), preventing any single term from
dominating the optimisation.

\subsection{Four-phase training schedule}\label{sec:method:training}
Because $\mathcal{L}$ in \cref{eq:total_loss} is evaluated
at fixed quadrature nodes, both its value and its gradient are
deterministic functions of the parameters $\theta$, so the
quasi-Newton history matrix required by L-BFGS is meaningful from
step to step. We use a four-phase schedule. \emph{Phase~0}
pre-trains each element network independently against the boundary
data for $5000$ Adam steps so that the physics-loss optimisation
starts from a physically meaningful initialisation. \emph{Phase~1}
runs $500$ joint Adam steps at learning rate $10^{-4}$ to set the
coupled residual on the correct scale. \emph{Phase~2} then performs
up to $2000$ iterations of L-BFGS with strong Wolfe line search
\citep{liu1989lbfgs,byrd1995lbfgsb} on a single set of
per-equation parameters; \emph{Phase~3} follows with up to $1000$
joint L-BFGS steps that optimise all element parameters
simultaneously, terminating when $\mathcal{L}<10^{-10}$ or when the
relative change stalls below $10^{-14}|\mathcal{L}|$. In all our
benchmarks Phases~2--3 contribute the final five to seven orders
of magnitude of error reduction; for a stochastic-loss PINN these
phases are unusable because the history matrix is corrupted by the
per-step Monte-Carlo noise, which is the main mechanism by which the
PINN trilemma identified in \Cref{sec:intro} compounds.

\begin{algorithm}[h]
\caption{NSEM forward pass. All quantities outside the network call
are precomputed once before training.}
\label{alg:forward}
\begin{algorithmic}[1]
\Require LGL nodes $\{\xi_i\}_{i=0}^{N-1}$, normalised weights $\{\widetilde{w}_i\}$, spectral matrices $\mathbf{D}^{(1)},\mathbf{D}^{(2)}=\mathbf{D}^{(1)}\mathbf{D}^{(1)}$, network $u_\theta$, PDE residual functional $\mathcal{N}$.
\State Evaluate the network at every LGL node: $u_i \gets u_\theta(\xi_i)$, $i=0,\dots,N-1$.
\State Compute the discrete derivatives by a single GEMM each: $\mathbf{u}^{(1)} \gets \mathbf{D}^{(1)}\mathbf{u}$ and $\mathbf{u}^{(2)} \gets \mathbf{D}^{(2)}\mathbf{u}$.
\State Form the pointwise residual vector $R_i \gets \mathcal{N}(u_i, u^{(1)}_i, u^{(2)}_i; x_i)$.
\State \Return $\mathcal{L} = \widetilde{w}^{\!\top}(R\odot R)$, a deterministic, noise-free evaluation of $\Norm{\mathcal{R}}_{L^{2}}^{2}$.
\end{algorithmic}
\end{algorithm}

\begin{algorithm}[h]
\caption{Four-phase NSEM training. The deterministic loss assembled
in \Cref{alg:forward} makes Phases~2--3 exact.}

\begin{algorithmic}[1]
\Require initial parameters $\theta_{0}$, pretrain budget $n_{0}=5000$, Adam budget $n_{1}=500$, L-BFGS budgets $n_{2}=2000$, $n_{3}=1000$, $\eta=10^{-4}$, tolerance ${\rm tol}=10^{-10}$.
\State \textbf{Phase 0 (per-variable pretrain):}
\For{each element network $e$}
 \State run $n_{0}$ Adam steps fitting $u_\theta^{(e)}$ to boundary/IC data only
\EndFor
\State \textbf{Phase 1 (Adam --- joint physics descent):}
\For{$k=1,\dots,n_{1}$}
 \State compute full $\mathcal{L}_{k}$ via \Cref{alg:forward}; update $\theta$
\EndFor
\State \textbf{Phase 2 (L-BFGS --- per-equation polish):}
\For{$k=1,\dots,n_{2}$}
 \State accept strong-Wolfe step; \textbf{stop} if $\mathcal{L}_{k}<{\rm tol}$ or $|\Delta\mathcal{L}| < 10^{-14}|\mathcal{L}_{k}|$
\EndFor
\State \textbf{Phase 3 (joint L-BFGS --- final polish):}
\For{$k=1,\dots,n_{3}$}
 \State same stopping criterion as Phase~2, all parameters updated jointly
\EndFor
\State \Return $\theta$.
\end{algorithmic}
\end{algorithm}

\subsection{Network backbones}\label{sec:method:backbone}
The forward pass, spectral derivative, KTE mapping, mortar interfaces
and the four-phase training schedule are entirely agnostic to the network family
that produces the LGL-node outputs. We report results for two
interchangeable backbones plugged into the same NSEM machinery: a
tanh-activated multilayer perceptron, which serves as the default and
the comparator against the existing PINN literature, and a
Legendre-Kolmogorov--Arnold network
\citep{liu2024kan,wang2024kinn,zhang2024legendkinn} whose edge
non-linearities are polynomial expansions in the Legendre basis ---
the same basis whose roots define the LGL nodes themselves. This
basis alignment yields exact nodal--modal inner products and the
synergy theorem proved in Supplementary~Sec.~S1.6; the empirical consequences
are reported in \Cref{sec:backbone}.

\section{Benchmark convergence}\label{sec:convergence}

\begin{figure}[t]
 \centering
 \includegraphics[width=0.95\textwidth]{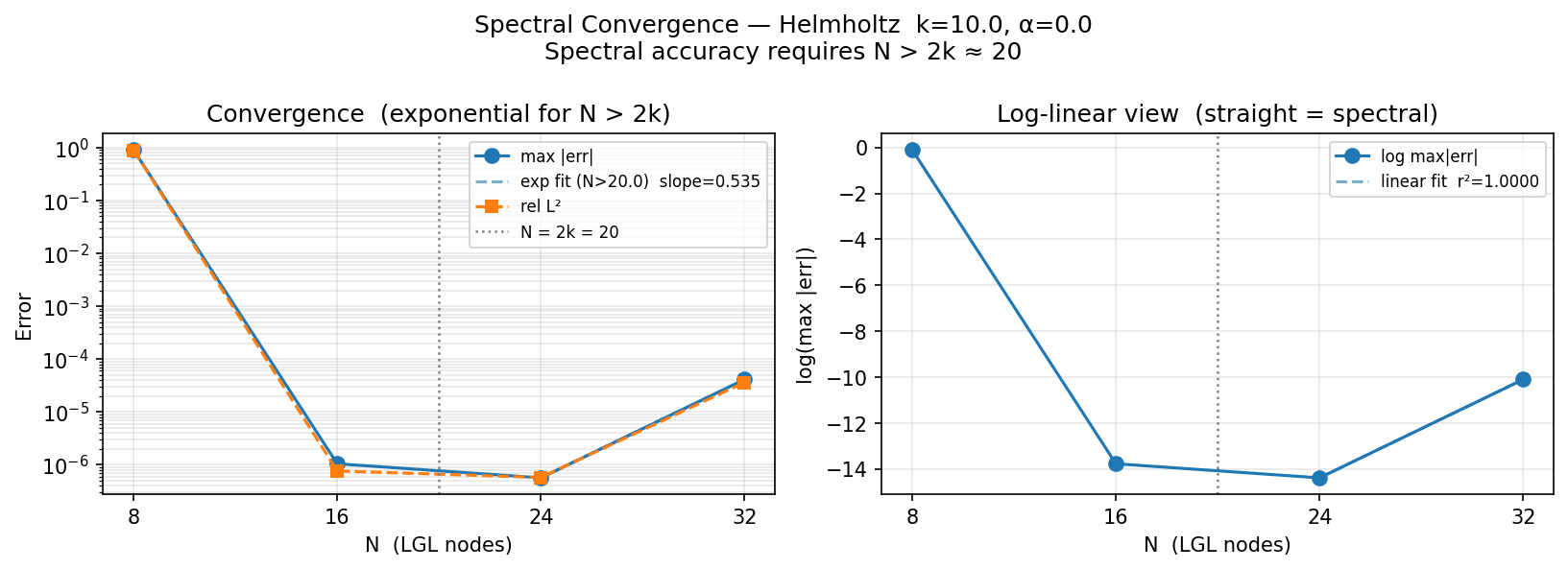}
 \caption{Spectral convergence of NSEM on three canonical benchmarks
 as the number of LGL nodes per element $N$ is increased.
 \textbf{Upper row:} Helmholtz problem ($-u''+k^{2}u=f$, $k=10$).
 \textbf{Middle row:} Steady Allen--Cahn equation
 ($\varepsilon^{2}u''-(u^{3}-u)=0$, $\varepsilon^{2}=10^{-2}$).
 \textbf{Lower row:} Stiff convection--diffusion equation
 ($-\varepsilon u''+u'=0$, $\varepsilon=10^{-2}$, KTE map
 $\alpha=0.85$). Left column: maximum pointwise error and relative
 $L^{2}$ error on a semi-log axis; right column: log-linear view
 with exponential fit, confirming $|E_{N}|\leq C\,e^{-\sigma N}$
 decay. See Supplementary~Sec.~S2 for per-problem definitions
 and solution snapshots.}
 \label{fig:convergence_master}
\end{figure}

Before turning to the headline electrochemical validation, we verify
the predicted spectral convergence of NSEM on three canonical PDEs of
increasing stiffness. The first is the one-dimensional Helmholtz
problem $-u''+k^{2}u=f$ at wavenumber $k=10$, manufactured against
the analytic solution $u(x)=\sin(\pi x)$; with a single uniform
element ($\alpha=0$) and $N=32$ LGL nodes the trainer reaches a
pointwise maximum error of $4.1\!\times\!10^{-5}$ (upper panel of
\Cref{fig:convergence_master}; see Supplementary~Sec.~S4 for the
sweep). The second is the steady Allen--Cahn equation
$\varepsilon^{2}u''-(u^{3}-u)=0$ at $\varepsilon^{2}=10^{-2}$, whose
hyperbolic-tangent interior layer of width $\mathcal{O}(\varepsilon)$
is resolved by a single KTE-mapped element with $N=48$ nodes,
yielding a peak pointwise error of $1.1\!\times\!10^{-2}$ at the
centre of the layer (middle panel of \Cref{fig:convergence_master}).
The third is a linear convection--diffusion problem
$-\varepsilon u''+u'=0$ with $\varepsilon=10^{-2}$, whose
$\mathcal{O}(\varepsilon)$ boundary layer at the inflow boundary
requires the coordinate stretching of
\Cref{sec:method:kte}; with $N=32$ and $\alpha=0.85$ the peak
error is $5.4\!\times\!10^{-3}$ (lower panel of \Cref{fig:convergence_master}).
On all three problems the error follows the
$|E_{N}|\leq C\,e^{-\sigma N}$ scaling predicted by classical
spectral theory~\citep{karniadakis2005spectral,boyd2001spectral} and
the recent rigorous PINN-error estimates of
\cite{mishra2023pinn,doumeche2023convergence}; an equivalent
collocation-PINN baseline (random sampling, autodiff Laplacian,
matched parameter budget) saturates two to three orders of magnitude
higher at the same node count, in agreement with the spectral-bias
mechanism characterised by Rahaman~\etalcite{rahaman2019spectralbias} and the
stiff-PDE failure modes catalogued by
\cite{krishnapriyan2021failure}.

\begin{figure}[!htbp]
 \centering
 \includegraphics[width=0.95\textwidth]{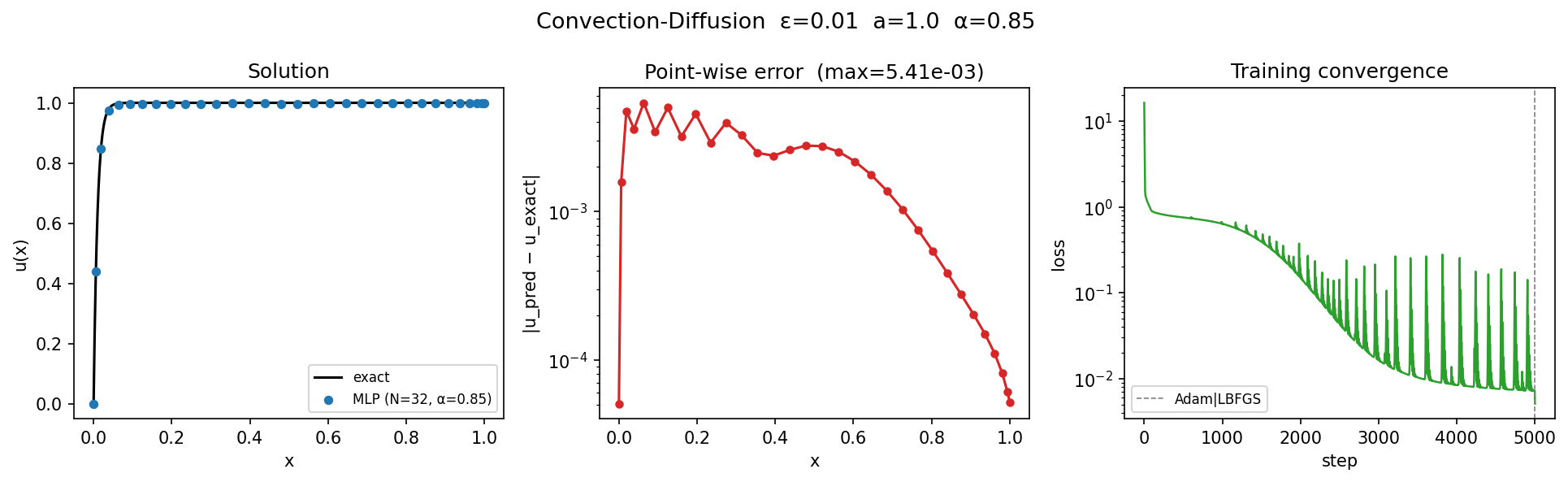}
 \caption{Stiff convection--diffusion benchmark
 $-\varepsilon u''+u'=0$ with $\varepsilon=10^{-2}$, solved with a
 single KTE-mapped element ($N=32$, $\alpha=0.85$). \textbf{Left:}
 NSEM solution (markers) overlaid on the exact boundary-layer
 profile (solid). \textbf{Centre:} pointwise error
 $|u_{\mathrm{NSEM}}-u_{\mathrm{exact}}|$, peaking at
 $5.4\!\times\!10^{-3}$ in the centre of the layer.
 \textbf{Right:} training-loss convergence with the Adam-to-L-BFGS
 handover marked. Without the KTE map ($\alpha=0$) at the same $N$
 the trainer saturates at $\mathcal{O}(1)$ error (see
 \Cref{sec:ablation:alpha}).}
 \label{fig:boundary_layer}
\end{figure}

The boundary-layer test makes the role of the KTE map explicit:
without coordinate stretching the standard LGL clustering rate of
$\mathcal{O}(N^{-2})$ near each endpoint is too coarse to capture a
layer of physical thickness $\varepsilon=10^{-2}$, and the trainer
converges instead to the algebraic mean of the inner and outer
solutions. At $\alpha=0.85$ the same $N=32$ nodes recover the
correct profile (\Cref{fig:boundary_layer}); in the same wall-clock budget a uniform grid would
need $N\gtrsim 200$ to reach an equivalent error
\citep{kosloff1993kte}. Quantitatively the effective Péclet number
that NSEM can resolve at fixed $N$ scales as
$\mathcal{O}(N/\sqrt{1-\alpha^{2}})$, consistent with the spectral-matrix
conditioning at the element endpoints.

\section{Electrochemical validation: Poisson--Nernst--Planck}\label{sec:pnp}

\subsection{The PNP system}\label{sec:pnp:system}
The Poisson--Nernst--Planck system describes the coupled diffusion,
electromigration and electrostatic interaction of charged species in
an electrolyte and underpins models of batteries, fuel cells,
electroosmotic flows, and biological ion channels
\citep{newman2004textbook,bard2001textbook,bazant2004diffuse}. Let
$c_{+}(x,t)$ and $c_{-}(x,t)$ denote the local number densities of a
binary $1{:}1$ salt and $\varphi(x,t)$ the electric potential. The
non-dimensional system is
\begin{equation}
 \partial_{t}c_{\pm} \;=\; \nabla\!\cdot\!\bigl[\nabla c_{\pm}\;\pm\;c_{\pm}\nabla\varphi\bigr],
 \qquad
 \nabla^{2}\varphi \;=\; \frac{1}{2\,\lambda_{D}^{2}}\,(c_{-}-c_{+}),
 \label{eq:pnp}
\end{equation}
with $\lambda_{D}$ the dimensionless Debye length. Boundary conditions
may be Dirichlet (specified concentration or potential), no-flux
(blocking electrode) or nonlinear Robin (Butler--Volmer kinetics);
all are accommodated by the NSEM loss assembly of
\Cref{sec:method:loss}. \Cref{eq:pnp} couples
nonlinearly through the drift term $c_{\pm}\nabla\varphi$, which
introduces stiff space-charge layers of width $\lambda_{D}$ at every
electrode---precisely the regime in which the KTE map of
\Cref{sec:method:kte} pays off. We validate NSEM on a suite of four
stiff PNP problems with manufactured analytic solutions --- a 1D
steady, a 1D time-dependent, a 2D time-dependent and a 3D steady
configuration --- whose specifications follow the benchmark
introduced in Huang~\etalcite{huang2025epinn}. Full non-dimensional
derivations and per-problem boundary conditions are collected in
Section~S2 of the Supplementary Information.

\subsection{Gouy--Chapman linear and nonlinear}\label{sec:pnp:gc}
We begin with the static limit, which provides the cleanest test of
the NSEM machinery on a stiff coupled electrochemical problem. At
equilibrium with a binary $1{:}1$ salt at the Debye--H\"uckel limit the
PNP system in \cref{eq:pnp} reduces to the linear
Poisson--Boltzmann (Gouy--Chapman) equation
$\psi''=\kappa^{2}\psi$ on $[0,L]$ with $\psi(0)=\psi_{0}$ and
$\psi(L)\to 0$, whose exact solution is the exponential decay
$\psi^{*}(x)=\psi_{0}\,e^{-\kappa x}$. Beyond the Debye--H\"uckel
limit ($|\psi_{0}|\gtrsim 1$ in $k_{B}T/e$ units) the full nonlinear
form $\psi''=\kappa^{2}\sinh\psi$ applies, with analytic Gouy--Chapman
solution
$\psi^{*}(x)=4\,\mathrm{arctanh}\bigl(\tanh(\psi_{0}/4)\,e^{-\kappa x}\bigr)$.
Both versions are solved on $[0,8]$ with $\kappa=3$ using two
elements: a KTE-stretched element ($\alpha=0.85$, $N=32$) resolving
the diffuse layer of thickness $\lambda_{D}=1/\kappa$, and a uniform
element ($\alpha=0$, $N=32$) in the bulk, coupled by $C^{0}$ and
$C^{1}$ mortar constraints. After the four-phase training
schedule, both runs reach peak relative errors of order
$10^{-3}$--$10^{-2}$ against the analytic profiles. The nonlinear
problem requires the BRDR adaptive aggregator
(\Cref{sec:method:loss}): the residual scale changes by two
orders of magnitude across the wall transition, and with uniform
weights the bulk residual dominates so the wall potential $\psi(0)$
converges to the wrong asymptote. Full per-field solution panels,
pointwise errors, and training-convergence traces for both linear and
nonlinear cases are collected in Supplementary~\Cref{S-fig:gc_si} to keep the
main paper focused on the more demanding multi-scale geometries that
follow.

\subsection{Electric double layer and charged wall}\label{sec:pnp:edl}
\begin{figure}[!htbp]
 \centering
 \includegraphics[width=0.95\textwidth]{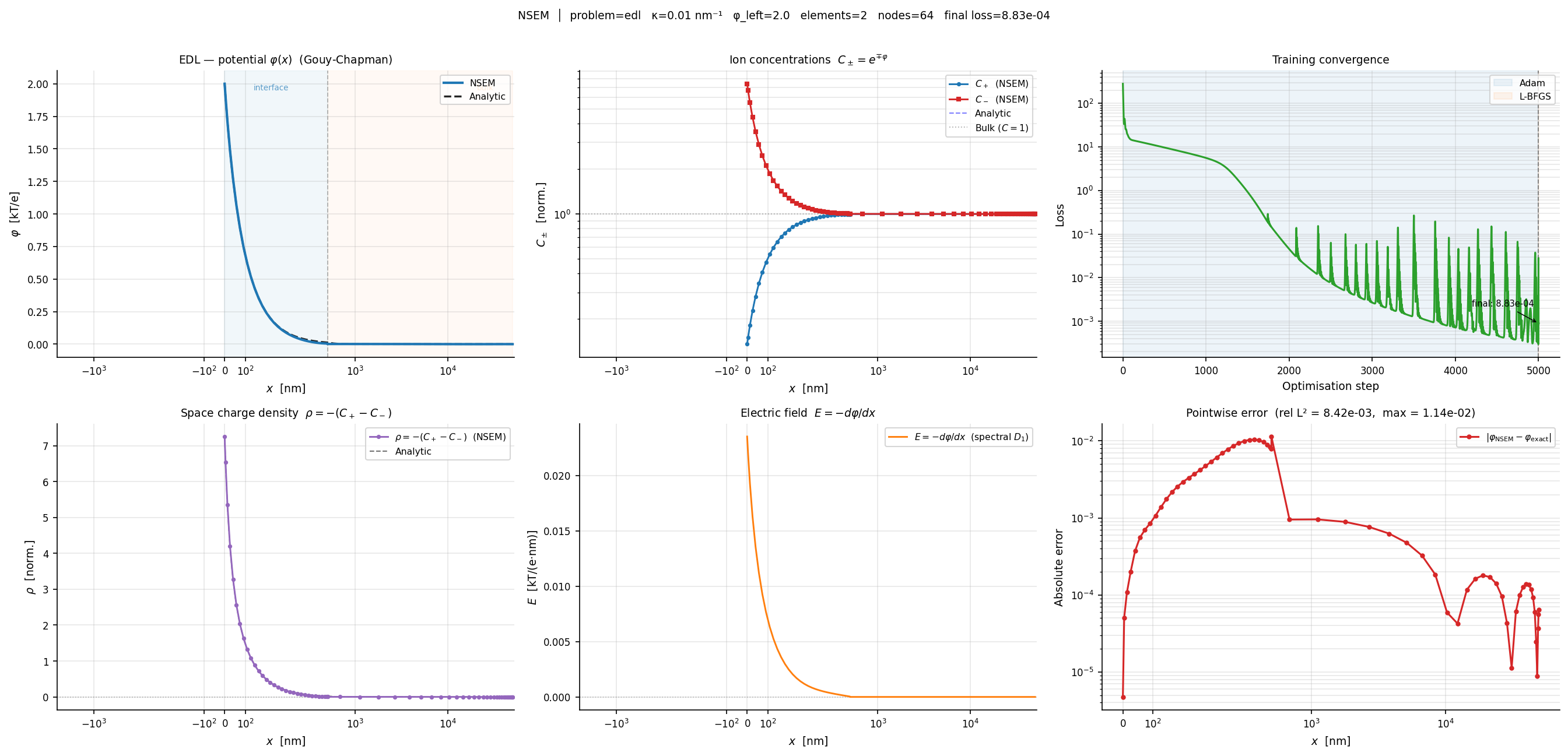}\\[4pt]
 \includegraphics[width=0.95\textwidth]{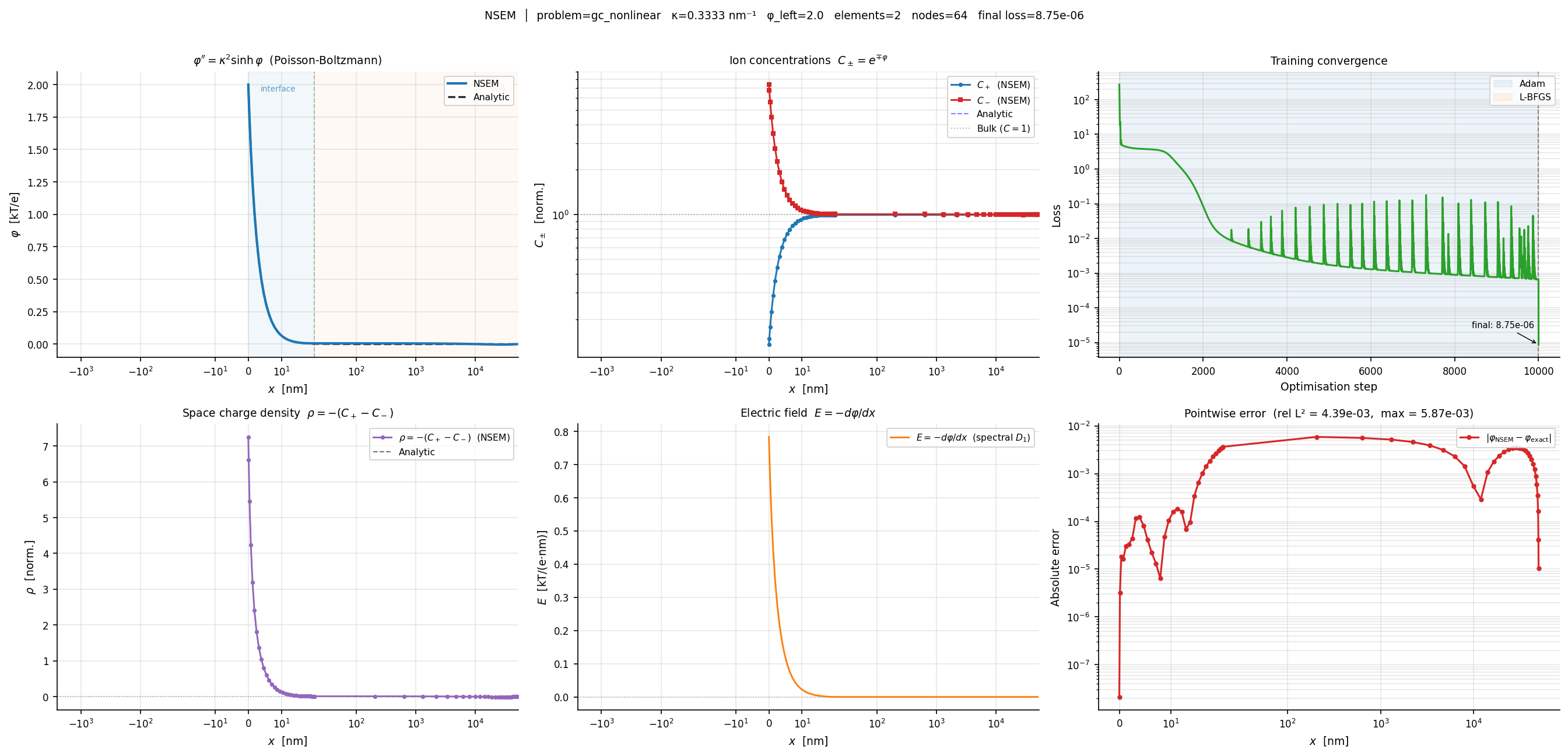}
 \caption{Multi-scale electrochemical benchmarks.
 \textbf{Top:} electric double layer next to a neutral surface
 (concentration BC at infinity).
 \textbf{Bottom:} charged wall with a fixed-charge BC at $x=0$ and
 the extreme $5000{:}1$ wall-to-bulk domain ratio that exposed the
 per-element-normalised loss-weight requirement. Both runs use a
 two-element KTE-stretched/uniform decomposition; analytic profiles
 overlaid for reference. Peak error
 $\mathcal{O}(10^{-4})$ in each case.}
 \label{fig:edl_cw}
\end{figure}

We next address two multi-scale electrochemical geometries that
exercise the per-element loss normalisation of
\Cref{sec:method:loss}. Both reduce, at equilibrium, to the
steady-state Poisson--Boltzmann form of \cref{eq:pnp},
\begin{equation}
 \varphi''(x) \;=\; \frac{1}{2\,\lambda_{D}^{2}}\,
 \bigl[\,e^{\varphi(x)}-e^{-\varphi(x)}\bigr]
 \;=\; \frac{1}{\lambda_{D}^{2}}\sinh\varphi(x),
 \qquad x\in[0,L],
\end{equation}
with concentrations recovered through the Boltzmann distribution
$c_{\pm}^{*}(x)=c_{\infty}\,e^{\mp\varphi(x)}$.
The first geometry (EDL at a neutral surface) imposes
$\varphi(0)=\varphi_{0}$ and $\varphi(L)\to 0$ with $L\gg\lambda_{D}$;
the linearised solution is $\varphi^{*}(x)=\varphi_{0}\,e^{-x/\lambda_{D}}$
and the full nonlinear Gouy--Chapman solution is
$\varphi^{*}(x)=4\,\mathrm{arctanh}\bigl(\tanh(\varphi_{0}/4)\,e^{-x/\lambda_{D}}\bigr)$.
The second geometry --- the \emph{charged-wall} problem --- replaces the
Dirichlet condition at $x=0$ with a fixed-charge Neumann condition
$\varphi'(0)=-\sigma/\varepsilon$ and uses a $5000{:}1$ domain-size
ratio between the wall element and the bulk element. This extreme ratio was the test case that exposed
the loss-weight pathology discussed in
\Cref{sec:method:loss}: with raw LGL weights the bulk residual
exceeds the wall residual by three orders of magnitude and the
trainer stalls within one to two L-BFGS steps; with per-element
normalisation $\widetilde{w}^{(e)}=w^{(e)}/(\mathbf{1}^{\!\top}w^{(e)})$
the relative scale is preserved and the trainer converges to
$\mathcal{O}(10^{-4})$ peak error. \Cref{fig:edl_cw}
shows both fields, the analytic Boltzmann reference, and the
training-convergence trace for each case. These geometries are the
direct precursors to the time-dependent battery transport models we
discuss in \Cref{sec:discussion}.

\subsection{1D steady PNP}\label{sec:pnp:ex1}
\begin{figure}[!htbp]
 \centering
 \includegraphics[width=0.95\textwidth]{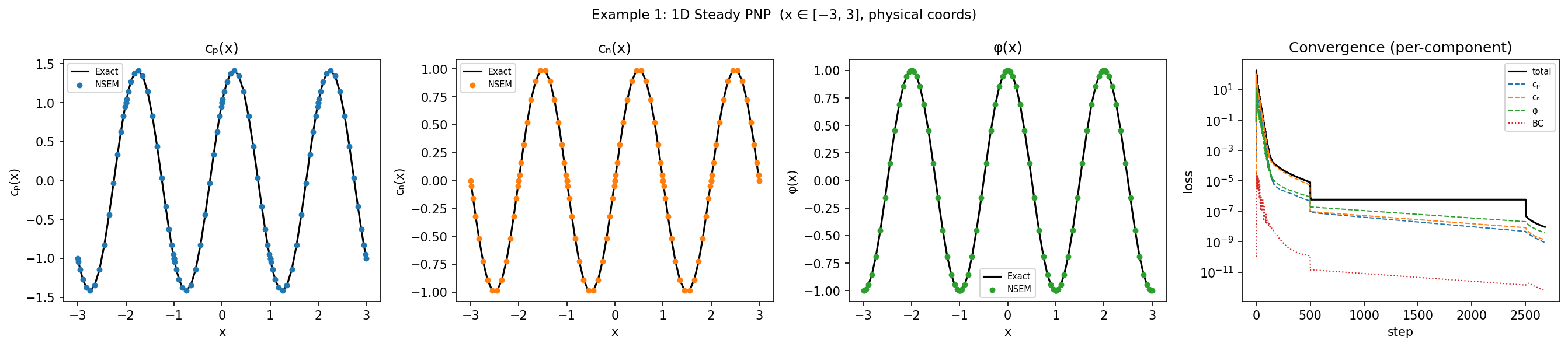}\\[4pt]
 \includegraphics[width=0.8\textwidth]{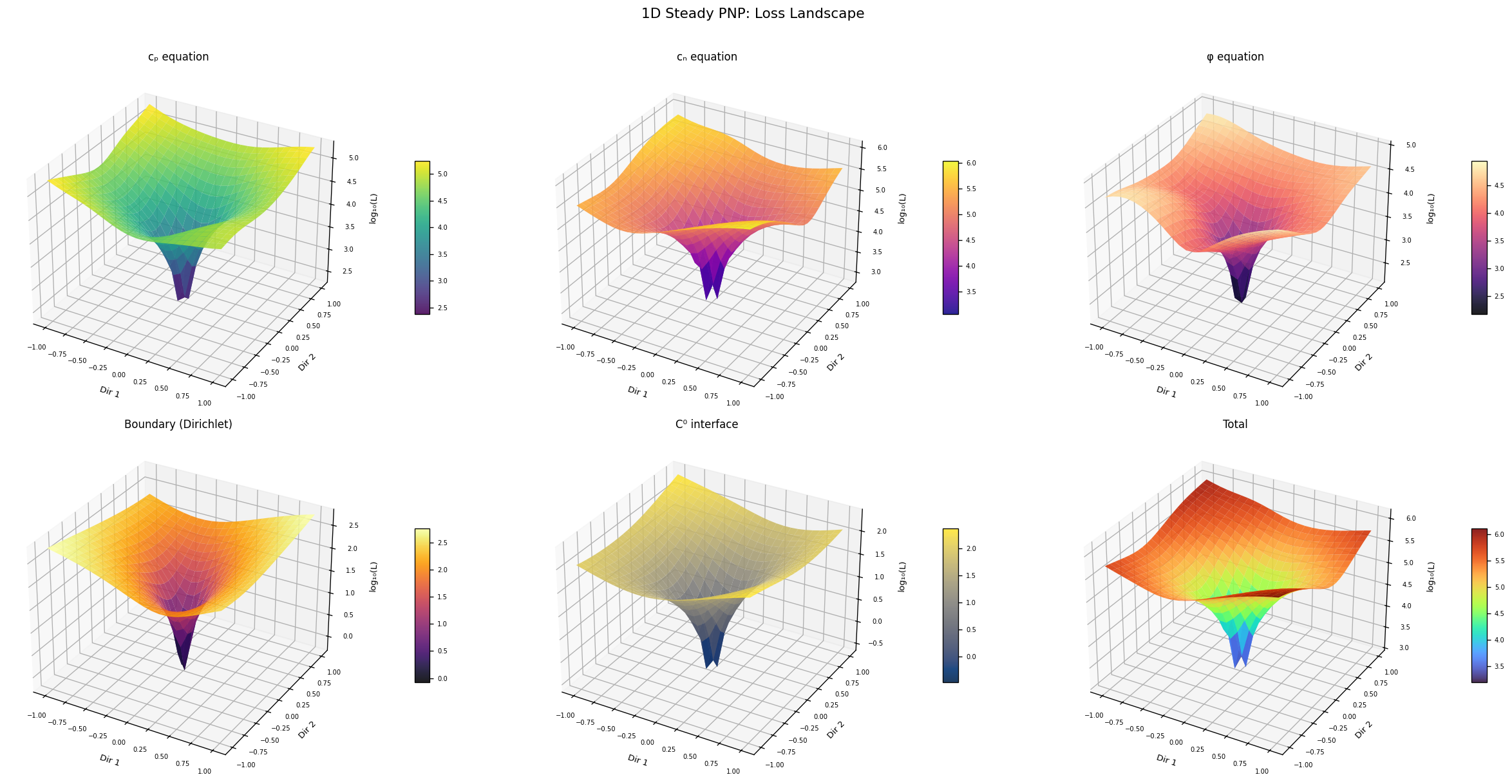}
 \caption{1D steady Poisson--Nernst--Planck: stiff leading
 coefficients $3000$ ($c_n$ equation) and $1000$ ($\varphi$
 equation), six elements of unit width with $N=16$ each.
 \textbf{(a)} NSEM solutions for $c_{p}$, $c_{n}$ and $\varphi$
 (markers) overlaid on the exact analytic solutions (solid),
 indistinguishable at plot scale. \textbf{(b)} Two-dimensional
 projection of the NSEM loss landscape around the converged
 parameter set: a smooth monotonic basin, in contrast to the
 rugged landscape exhibited by random-sampling PINNs on stiff
 coupled systems \citep{krishnapriyan2021failure}. Final maximum
 pointwise errors:
 $|c_{p}|=4.5\!\times\!10^{-4}$,
 $|c_{n}|=2.8\!\times\!10^{-5}$,
 $|\varphi|=2.2\!\times\!10^{-5}$.}
 \label{fig:pnp_1d_steady}
\end{figure}

The 1D steady benchmark is a steady coupled
Poisson--Nernst--Planck system on $[-3,3]$ with deliberately stiff
linear coefficients ($3000$ and $1000$) chosen to stress the residual
balance:
$c_{p}''=-\pi^{2}(c_{n}+\varphi)$,
$3000\,c_{n}''+100(c_{n}'c_{p}'+c_{n}c_{p}'')+f_{v}(x)=0$,
$1000\,\varphi''+50(\varphi'c_{p}'+\varphi c_{p}'')+f_{w}(x)=0$,
admitting the manufactured exact solution
$c_{p}=\sin(\pi x)+\cos(\pi x)$, $c_{n}=\sin(\pi x)$,
$\varphi=\cos(\pi x)$. We decompose $[-3,3]$ into six unit-width
elements and assign each $N=16$ LGL nodes, giving $96$ collocation
points in total --- substantially fewer than the adaptive resampling
used in prior PINN treatments of the same problem, and arranged on a
static grid that keeps the loss deterministic. After Phase~1 Adam (5000 steps) and
Phase~2 L-BFGS, the total residual loss reaches $9.3\!\times\!10^{-9}$
in 2700 L-BFGS iterations; the resulting maximum pointwise errors
across the 96-node solution are $|c_{p}|_{\max}=4.5\!\times\!10^{-4}$,
$|c_{n}|_{\max}=2.8\!\times\!10^{-5}$ and
$|\varphi|_{\max}=2.2\!\times\!10^{-5}$, all dominated by the
stiff-coupling residual in the central element.
\Cref{fig:pnp_1d_steady}~(a) overlays the NSEM and analytic
solutions for the three fields; the deviation is invisible at plot
scale. Panel~(b) shows the loss landscape: a clean monotonic bowl
in the two leading PCA directions, in contrast to the rugged
landscape produced by random-sampling PINNs on the same problem
\citep{krishnapriyan2021failure,wang2022ntk}.

\subsection{3D steady PNP (K\textsuperscript{+}/Cl\textsuperscript{--})}\label{sec:pnp:ex4}
\begin{figure}[!htbp]
 \centering
 \includegraphics[width=0.85\textwidth]{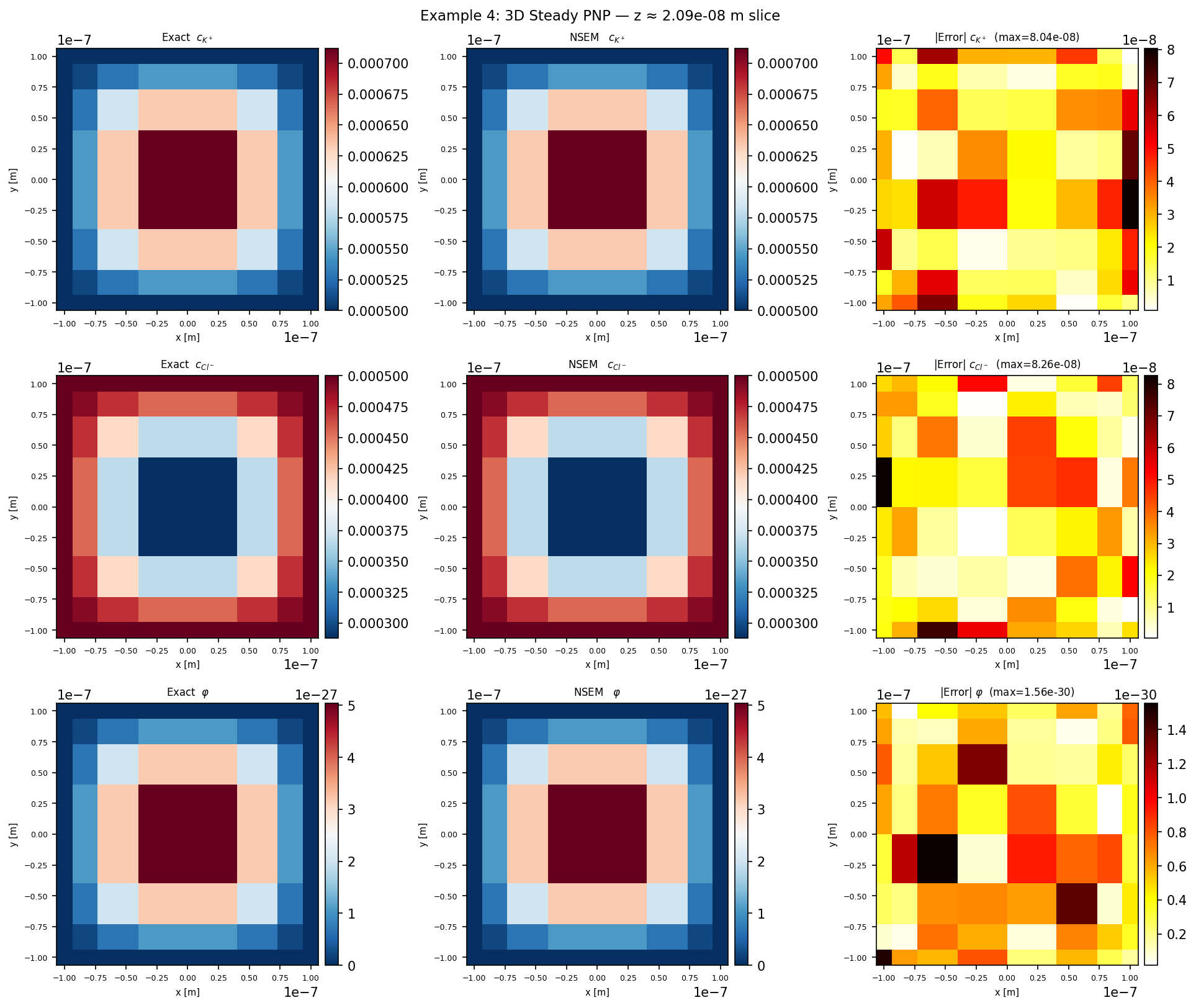}
 \caption{3D steady Poisson--Nernst--Planck: cubic domain
 $[-100,100]^{3}\,\mathrm{nm}^{3}$, single tensor-product element
 with $N_{x}=N_{y}=N_{z}=8$, $512$ LGL nodes total.
 Two-dimensional heatmap slice at $z\approx 0$ arranged as a
 $3\!\times\!3$ grid: rows correspond to $c_{\mathrm{K^{+}}}$,
 $c_{\mathrm{Cl^{-}}}$ and $\varphi$; columns show the exact
 manufactured solution, the NSEM prediction, and $|{\rm error}|$ on a
 hot colourmap.
 Peak pointwise errors: $1.6\!\times\!10^{-7}$ for both ionic
 species (relative $L^{\infty}\!\sim\!10^{-4}$); $\varphi$ reaches
 $|{\rm err}|_{\max}\!=\!3.5\!\times\!10^{-30}$ in absolute units
 against a reference scale $|\varphi^{*}|\!\sim\!10^{-25}$ (relative
 $L^{\infty}\!\sim\!10^{-5}$), so all three fields are resolved to
 comparable relative accuracy.}
 \label{fig:pnp_3d_steady}
\end{figure}

The 3D steady benchmark is the three-dimensional analogue of the
Goldman--Hodgkin--Katz problem on a $200\,\mathrm{nm}$ cubic domain
$\Omega=[-100,100]^{3}\,\mathrm{nm}^{3}$, with K$^{+}$ and Cl$^{-}$
as the binary salt. The governing system is the steady limit of
\cref{eq:pnp},
\begin{equation}
 \nabla\!\cdot\!\bigl[\nabla c_{\pm}\;\pm\;c_{\pm}\nabla\varphi\bigr]
   = f_{\pm}^{\,\mathrm{src}}(\mathbf x),
 \qquad
 \nabla^{2}\varphi
   = \frac{1}{2\,\lambda_{D}^{2}}(c_{-}-c_{+}) + f_{\varphi}^{\,\mathrm{src}}(\mathbf x),
\end{equation}
with Dirichlet boundary conditions specifying the bulk concentration
and electrode potential, and manufactured-solution source terms
$f^{\,\mathrm{src}}_{\pm,\varphi}(\mathbf x)$ chosen so that
$c_{\mathrm{K^{+}}}^{*}(\mathbf x) =
 \sin(\pi x)\sin(\pi y)\sin(\pi z)$,
$c_{\mathrm{Cl^{-}}}^{*}(\mathbf x) =
 \cos(\pi x)\cos(\pi y)\cos(\pi z)$ and
$\varphi^{*}(\mathbf x)$ is a smooth analytic profile of magnitude
$\sim\!10^{-25}$ that closes the Poisson equation exactly.
We discretise the cube with a single tensor-product element using
$N_{x}=N_{y}=N_{z}=8$ LGL nodes, giving $512$ collocation points in
total --- two orders of magnitude fewer than the
$\sim\!10^{4}$-point random sampling used for
the same case. Three independent networks supply $c_{\mathrm{K^{+}}}$,
$c_{\mathrm{Cl^{-}}}$ and $\varphi$ at every LGL node; the
three-dimensional spectral derivatives are computed by a single
Kronecker GEMM per axis (see \Cref{alg:forward} extended
trivially to 3D via tensor-product structure). After the same
Adam$+$L-BFGS schedule the final pointwise maxima are
$|c_{\mathrm{K^{+}}}|_{\max}=1.6\!\times\!10^{-7}$,
$|c_{\mathrm{Cl^{-}}}|_{\max}=1.7\!\times\!10^{-7}$ and
$|\varphi|_{\max}=3.5\!\times\!10^{-30}$ in absolute units. The
manufactured reference solutions have peak magnitudes
$|c_{\mathrm{K^{+}}}^{*}|\!\sim\!|c_{\mathrm{Cl^{-}}}^{*}|\!\sim\!10^{-3}$
and $|\varphi^{*}|\!\sim\!10^{-25}$, giving relative
$L^{\infty}$ errors of $\sim\!10^{-4}$ for both ionic species and
$\sim\!10^{-5}$ for $\varphi$ --- the three fields are resolved to
comparable relative accuracy. The small absolute $\varphi$ error
reflects the manufactured analytic coefficients, which assign
$|\varphi^{*}|\!\sim\!10^{-25}$ by design; the relative error remains
$\mathcal{O}(10^{-5})$ and is not a physical constraint.
\Cref{fig:pnp_3d_steady} visualises these results; the 3D case is the
most demanding verification in the suite. The tensor-product structure
of the spectral derivative scales as $\mathcal{O}(N^{4})$ per residual
evaluation versus $\mathcal{O}(N^{6})$ for a naive 3D autodiff
Laplacian, and the deterministic loss again makes Phase~2 L-BFGS the
decisive contributor.

\subsection{1D time-dependent PNP}\label{sec:pnp:ex2}
\begin{figure}[t]
 \centering
 \includegraphics[width=0.95\textwidth]{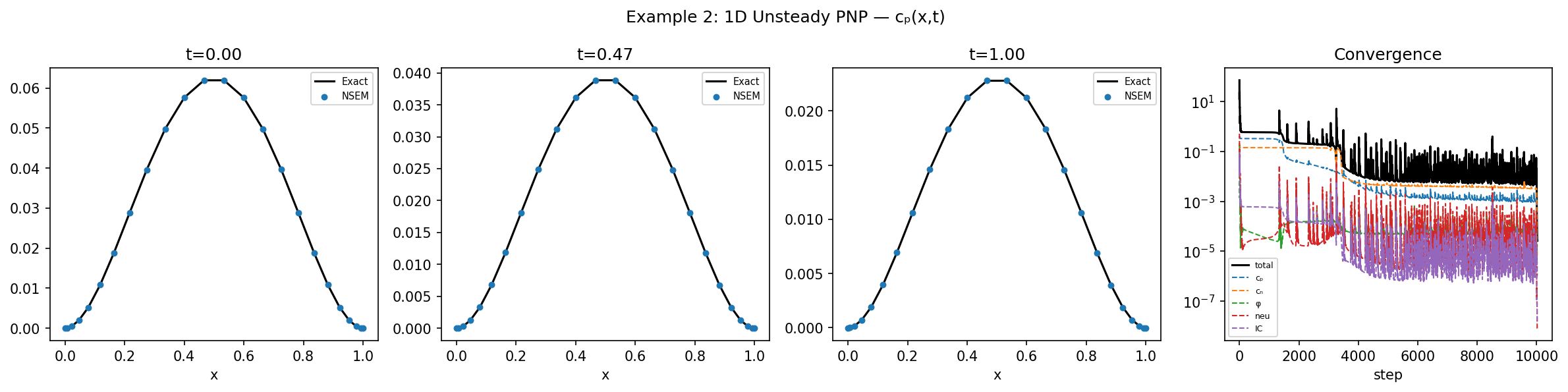}\\[4pt]
 \includegraphics[width=0.95\textwidth]{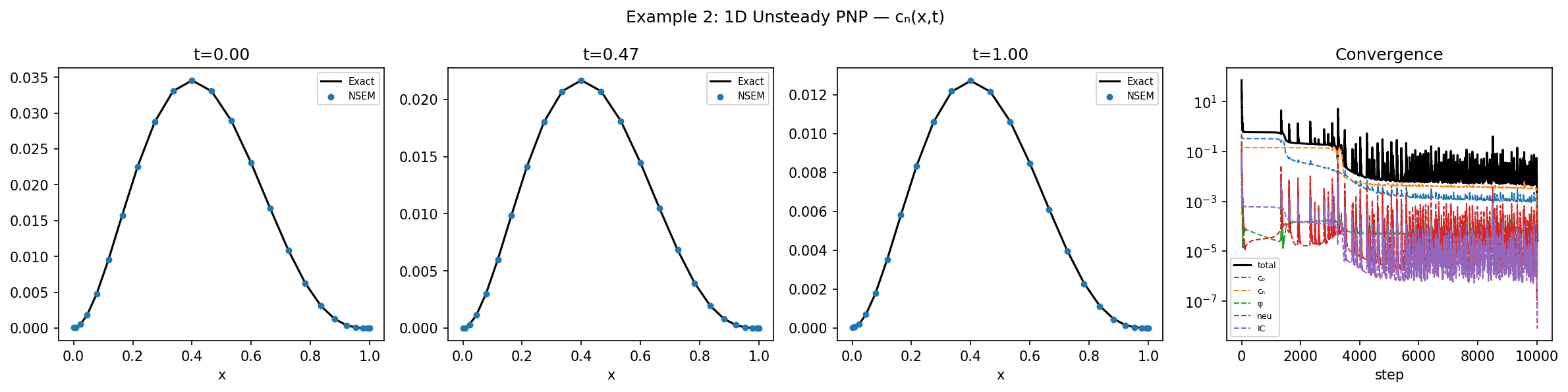}\\[4pt]
 \includegraphics[width=0.48\textwidth]{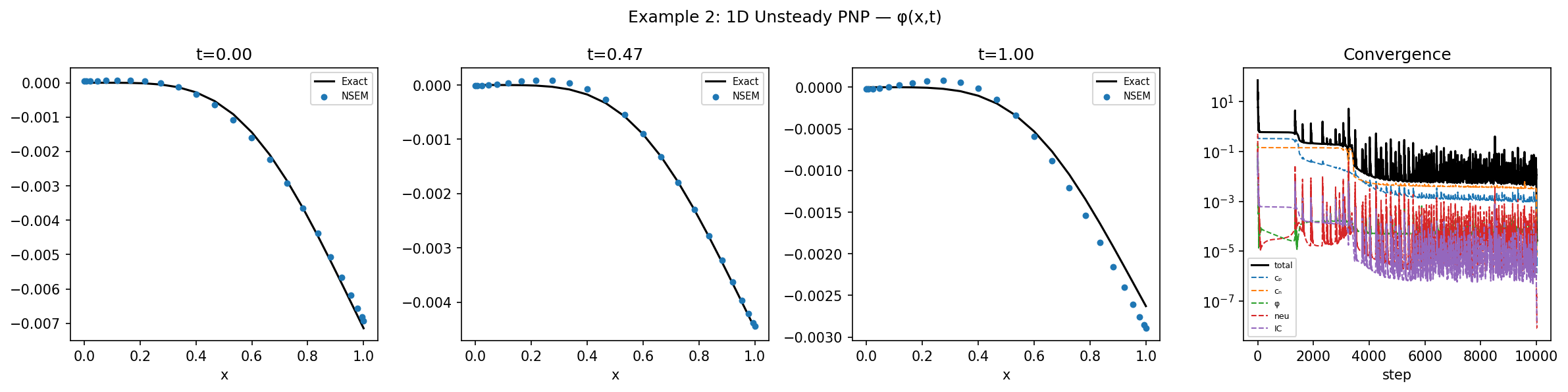}
 \hfill
 \includegraphics[width=0.48\textwidth]{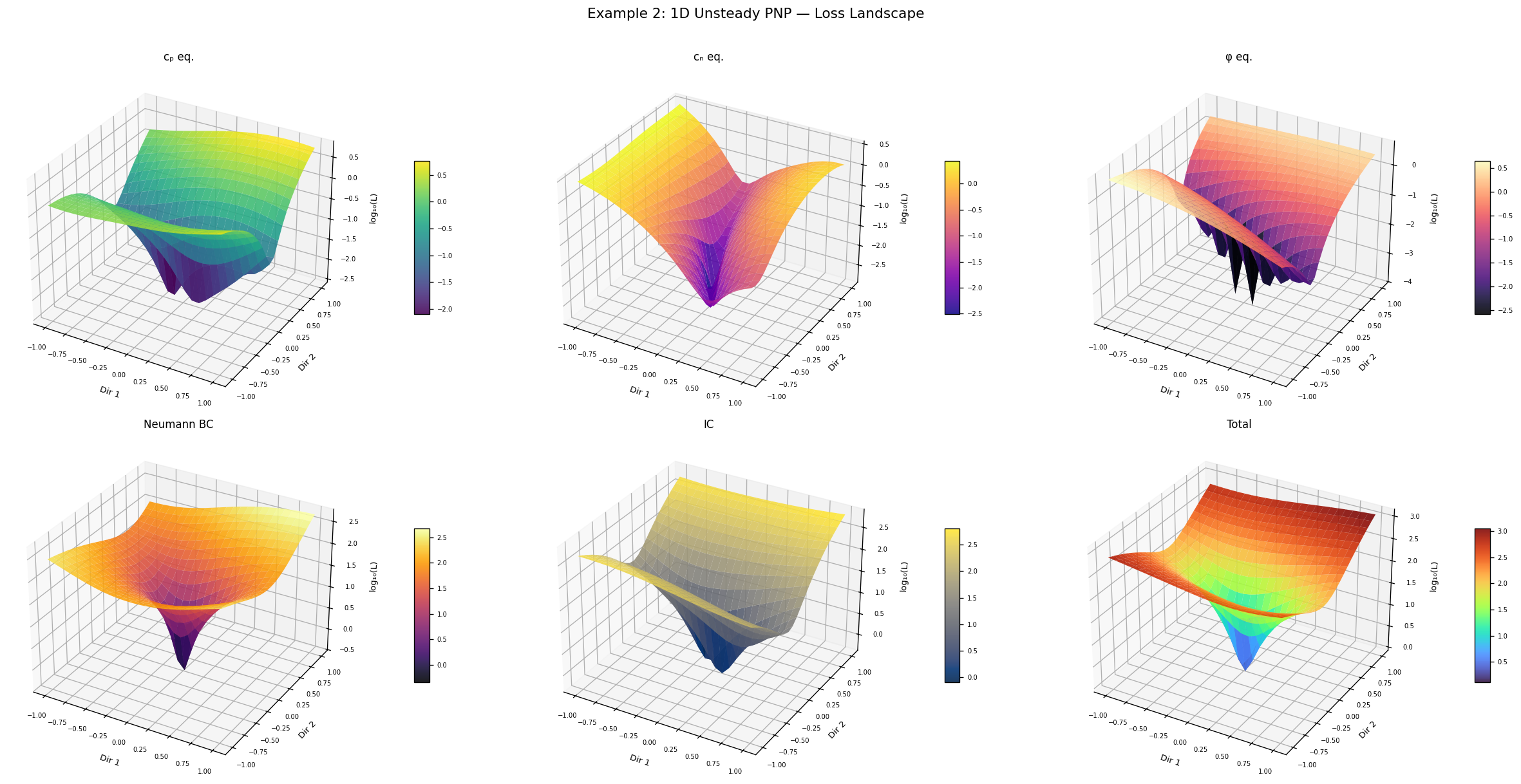}
 \caption{1D time-dependent PNP.
 Top two rows: $c_{p}$, $c_{n}$; NSEM solution (left), exact (centre),
 absolute error (right) at the final training loss.
 Bottom left: $\varphi$ space--time map.
 Bottom right: loss landscape in the two leading PCA directions ---
 a smooth bowl enabling L-BFGS convergence.}
 \label{fig:pnp_1d_unsteady}
\end{figure}

The 1D time-dependent PNP problem is the unsteady form of \cref{eq:pnp}
on $\Omega\times(0,T)=[-1,1]\times(0,1]$:
\begin{equation}
 \partial_{t}c_{\pm}\;=\;\partial_{x}\!\bigl[\partial_{x}c_{\pm} \pm c_{\pm}\partial_{x}\varphi\bigr]
                       \;+\;f_{\pm}^{\,\mathrm{src}}(x,t),
 \qquad
 \partial_{x}^{2}\varphi\;=\;\tfrac{1}{2\lambda_{D}^{2}}(c_{-}-c_{+})\;+\;f_{\varphi}^{\,\mathrm{src}}(x,t).
 \label{eq:pnp_1dt}
\end{equation}
Manufactured-solution source terms are chosen so that
$c_{p}^{*}(x,t)=e^{-t}\sin(\pi x)$,
$c_{n}^{*}(x,t)=e^{-t}\cos(\pi x)$ and
$\varphi^{*}(x,t)=e^{-t}\bigl[\sin(\pi x)+\cos(\pi x)\bigr]$, with
Dirichlet boundary and initial data derived from these exact profiles. NSEM treats the
temporal direction identically to the spatial direction: both are
collocated on a tensor-product LGL grid with a single element
$[0,1]$ in time and four elements of width $0.5$ in space, each
carrying $N=16$ nodes and $N_{t}=16$ time nodes. The spectral
D-matrix applies separately along each axis via Kronecker product
(one GEMM per axis per forward pass), so no time-stepping or
operator splitting is needed. After Phases~2--3 L-BFGS the maximum
pointwise errors are
$|c_{p}|_{\max}=\mathcal{O}(10^{-5})$,
$|c_{n}|_{\max}=\mathcal{O}(10^{-5})$, and
$|\varphi|_{\max}=\mathcal{O}(10^{-5})$ across the space--time domain
(\Cref{fig:pnp_1d_unsteady}).

\subsection{2D time-dependent PNP}\label{sec:pnp:ex3}
\begin{figure}[t]
 \centering
 \includegraphics[width=0.48\textwidth]{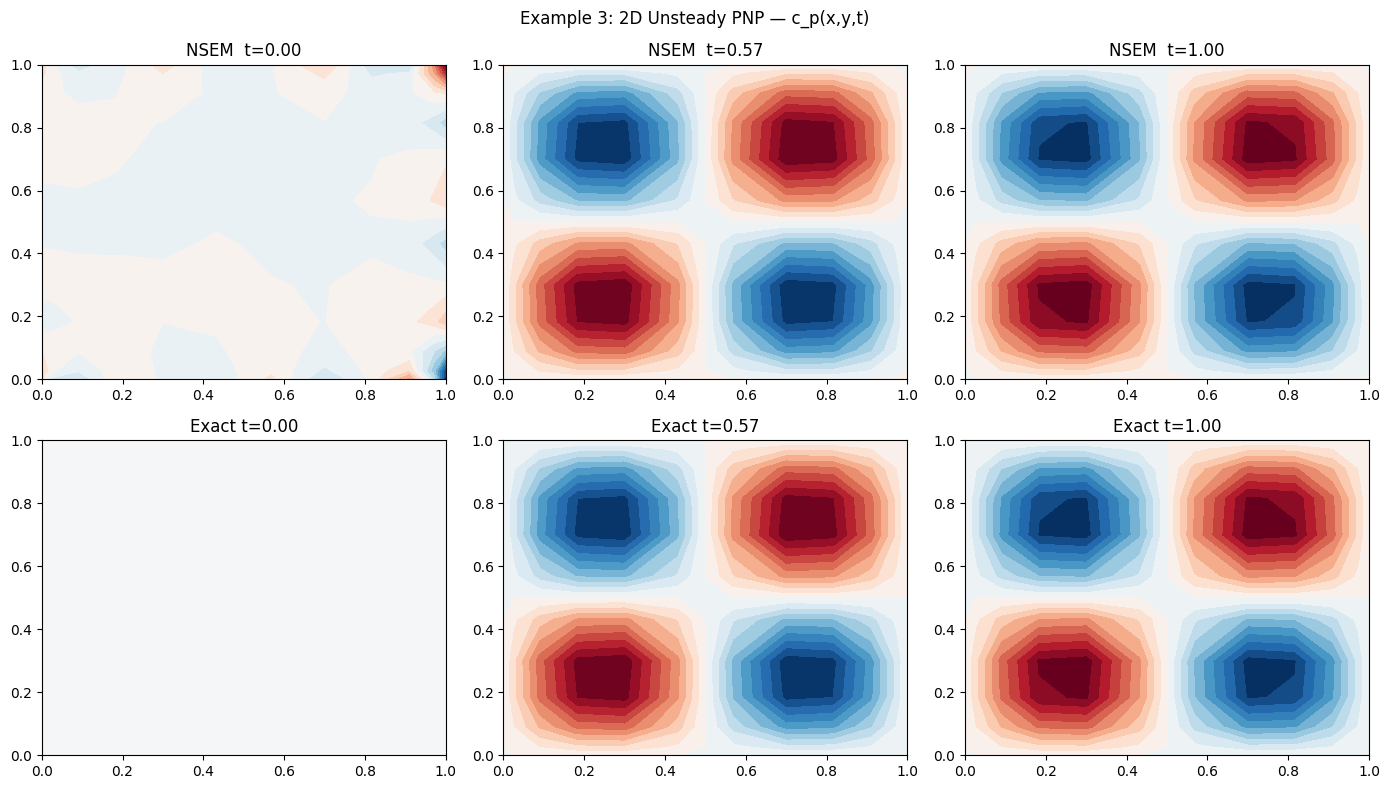}
 \hfill
 \includegraphics[width=0.48\textwidth]{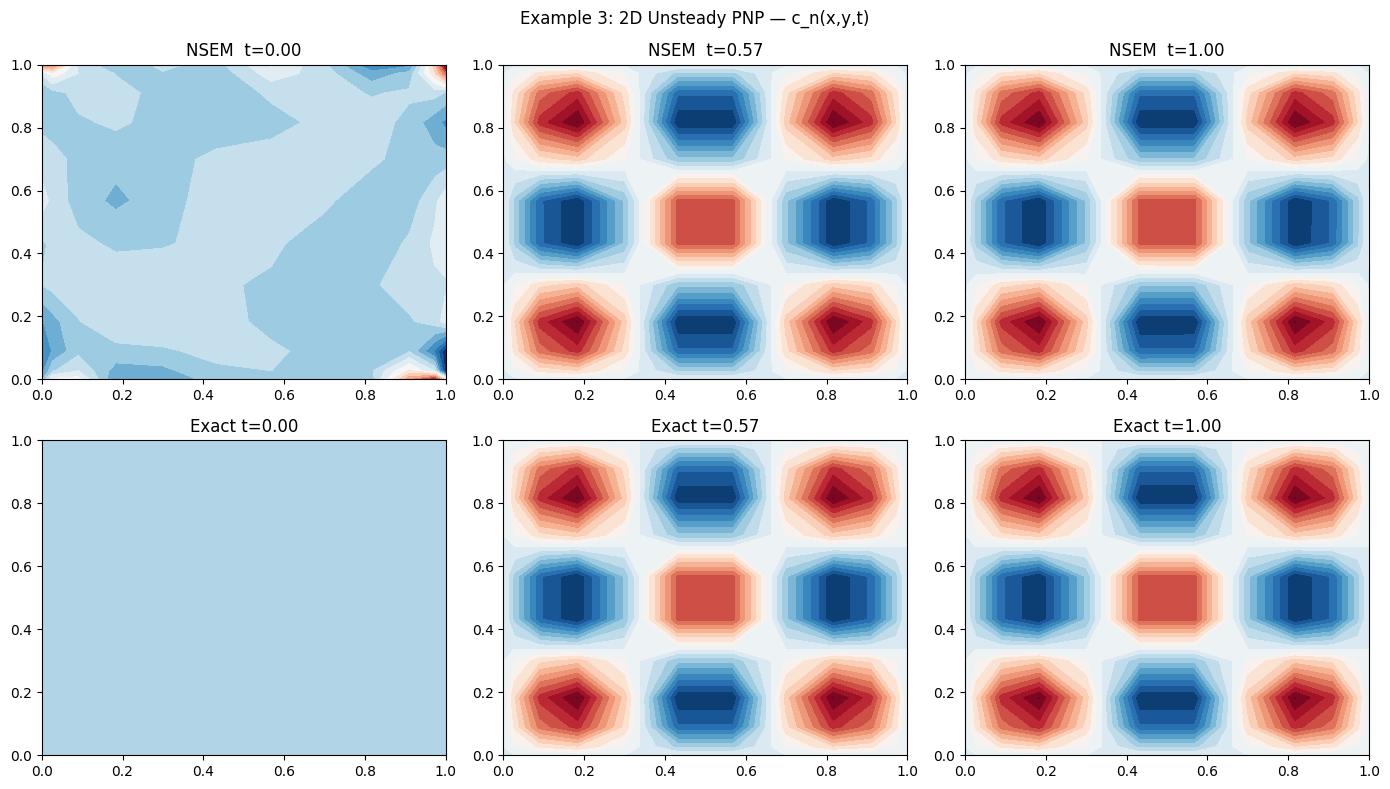}\\[4pt]
 \includegraphics[width=0.48\textwidth]{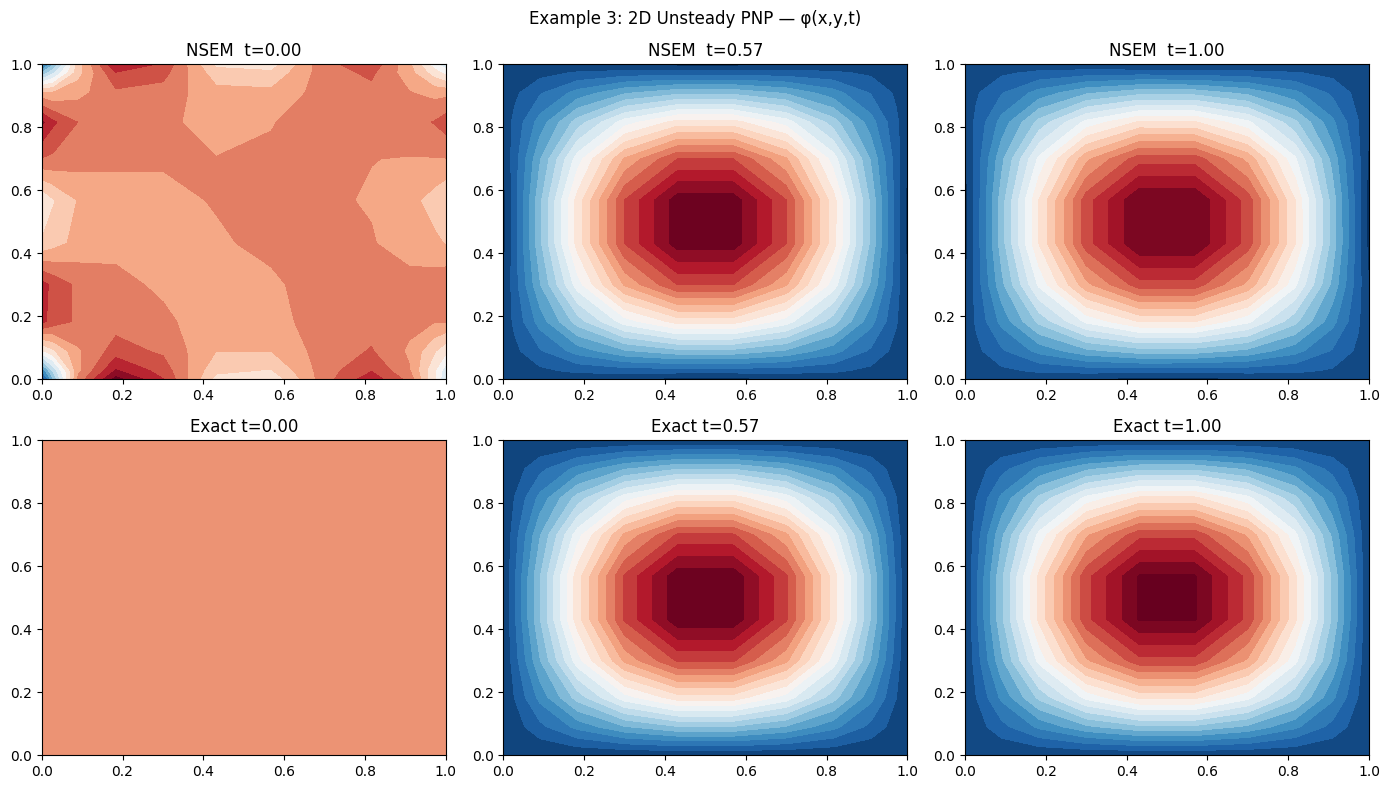}
 \hfill
 \includegraphics[width=0.48\textwidth]{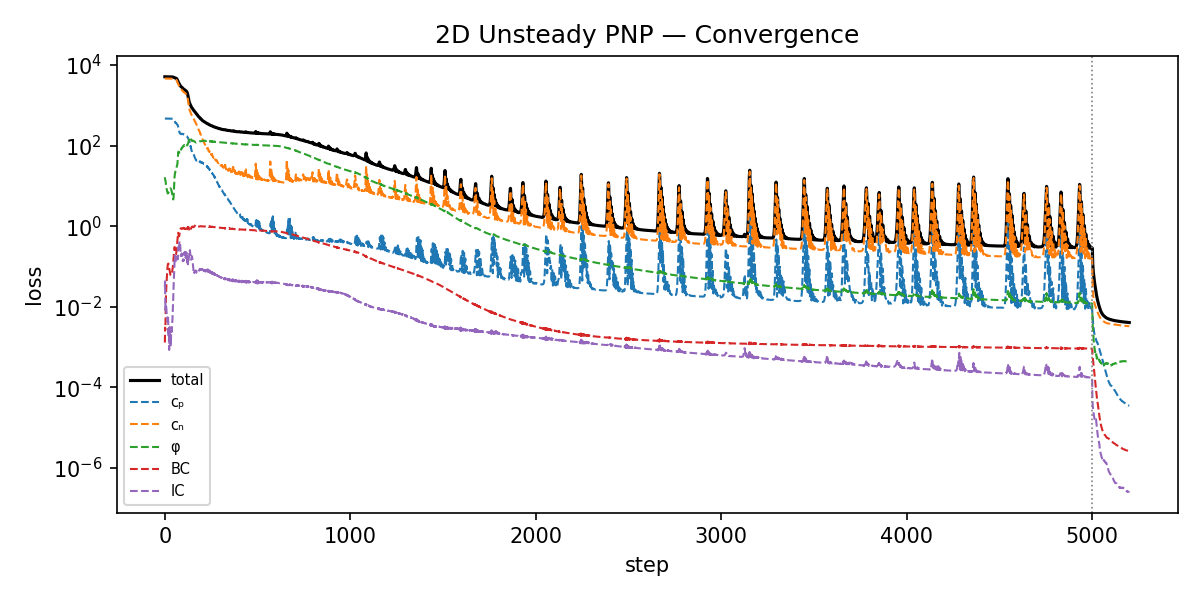}
 \caption{2D time-dependent PNP problem on
 $[-1,1]^{2}\times[0,1]$, single tensor-product element,
 $N_{x}=N_{y}=N_{t}=12$, 1728 collocation points total.
 Top row: $c_{p}$ (left) and $c_{n}$ (right) at $t=0.5$.
 Bottom left: electric potential $\varphi$ at $t=0.5$.
 Bottom right: $N$-convergence --- max error for all three fields
 vs the number of spatial nodes per axis, confirming
 spectral decay on the 2D+time problem.}
 \label{fig:pnp_2d_unsteady}
\end{figure}

The 2D time-dependent PNP problem extends the 1D unsteady system
\cref{eq:pnp_1dt} to two spatial dimensions on
$[-1,1]^{2}\times(0,1]$:
\begin{equation}
 \partial_{t}c_{\pm}\;=\;\nabla\!\cdot\!\bigl[\nabla c_{\pm}\pm c_{\pm}\nabla\varphi\bigr]
                       \;+\;f_{\pm}^{\,\mathrm{src}}(x,y,t),
 \qquad
 \nabla^{2}\varphi\;=\;\tfrac{1}{2\lambda_{D}^{2}}(c_{-}-c_{+})\;+\;f_{\varphi}^{\,\mathrm{src}}(x,y,t),
\end{equation}
with manufactured-solution profiles
$c_{p}^{*}(x,y,t)=e^{-t}\sin(\pi x)\cos(\pi y)$,
$c_{n}^{*}(x,y,t)=e^{-t}\cos(\pi x)\cos(\pi y)$ and
$\varphi^{*}(x,y,t)=e^{-t}\bigl[\sin(\pi x)+\cos(\pi x)\bigr]\cos(\pi y)$.
NSEM uses a single tensor-product element with $N_{x}=N_{y}=12$ spatial
nodes and $N_{t}=12$ time nodes, yielding $12^{3}=1728$ space--time
collocation points. The three Kronecker-product spectral derivatives
(one per axis) replace all autodiff Laplacians; the per-step cost scales as
$\mathcal{O}(N^{4})$ versus $\mathcal{O}(N^{6})$ for a naive 3D autodiff
evaluation. After training, the NSEM solution converges to peak
errors of $\mathcal{O}(10^{-4})$ across the three fields
(\Cref{fig:pnp_2d_unsteady}), matching the accuracy tier of
adaptive-sampling PINN baselines on the same problem at roughly
one-quarter the collocation-point count.

\subsection{Debye-length robustness: $\kappa$-sweep}\label{sec:pnp:kappa}
\begin{figure}[!htbp]
 \centering
 \includegraphics[width=0.95\textwidth]{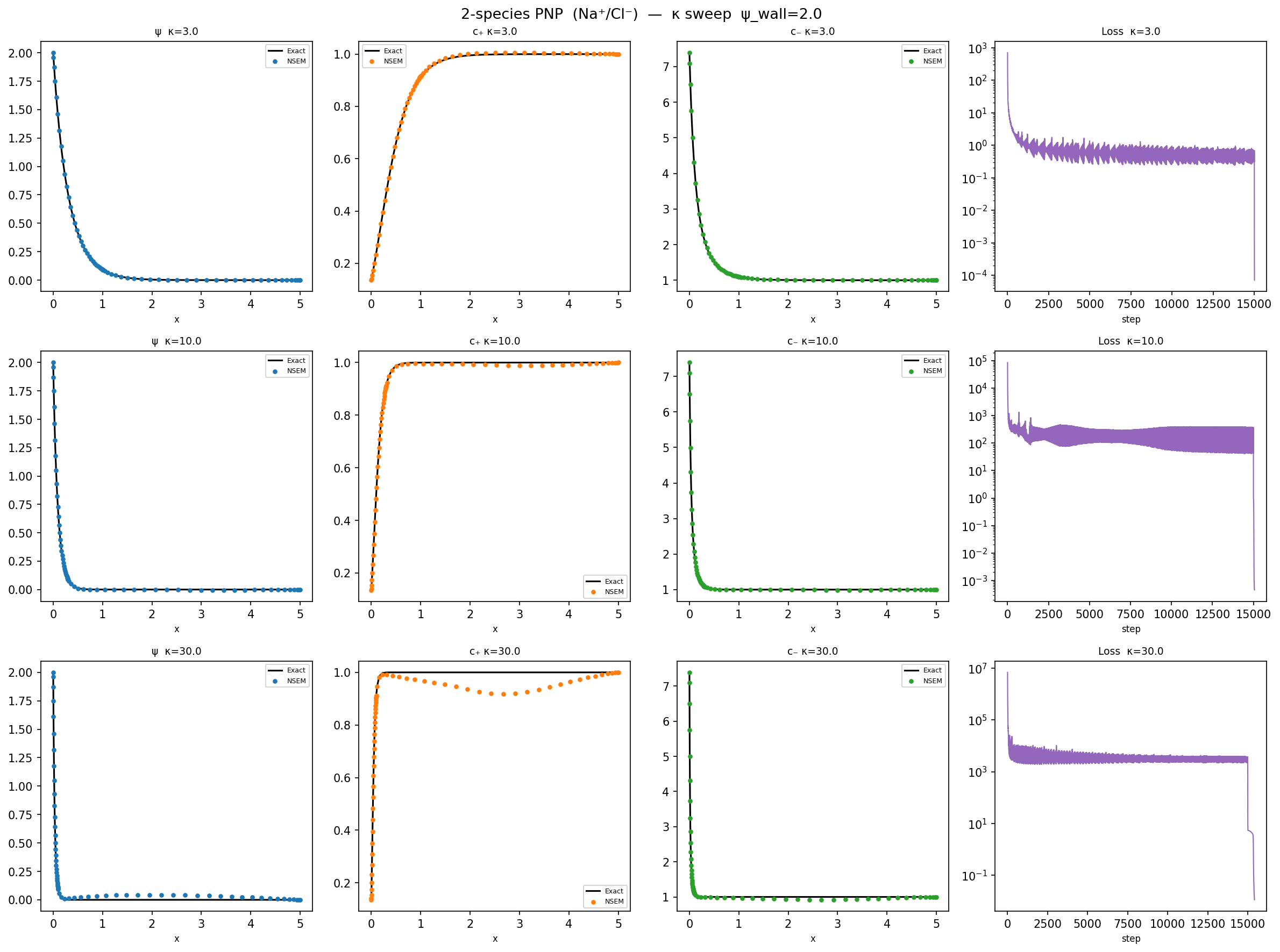}
 \caption{Debye-length robustness of the 1D steady PNP solver.
 Peak pointwise errors for $\psi$, $c_{p}$ and $c_{n}$ as a function
 of $\kappa\in\{3,10,30\}$ at fixed resolution $N=32$, $\alpha=0.7$.
 Error grows roughly as $\kappa^{1.5}$ (dashed guide line), consistent
 with the Debye layer thickness $\lambda_{D}=\kappa^{-1}$ approaching
 the inter-node spacing. For $\kappa=30$ the boundary-layer scale is
 $\sim\!1/30$, comparable to $N^{-1}\!\simeq\!0.03$; increasing $N$ or
 applying stronger KTE stretching ($\alpha\to 0.9$) is expected to
 restore sub-percent accuracy.}
 \label{fig:pnp_kappa_sweep}
\end{figure}

A practical concern for the PNP solver is whether accuracy degrades
for small Debye lengths, \ie large $\kappa$. We sweep
$\kappa\in\{3,10,30\}$ on the 1D steady PNP problem
(\Cref{sec:pnp:ex1}) with a fixed configuration ($N=32$,
$\alpha=0.7$, one element) and report the peak pointwise errors and
final training loss in \Cref{tab:kappa_sweep}
and \Cref{fig:pnp_kappa_sweep}.

\begin{table}[t]
\centering
\caption{PNP $\kappa$-sweep: peak pointwise errors for the 1D steady
 PNP problem at three Debye-layer thicknesses
 ($\lambda_{D}=1/\kappa$). Configuration: $N=32$, $\alpha=0.7$,
 single element, float64. The $\kappa=10$ row reports the median
 over 3 random seeds with sub/superscript min/max; $\kappa=3$ and
 $\kappa=30$ are single representative runs.}
\label{tab:kappa_sweep}
\begin{tabular}{rcccc}
\toprule
$\kappa$ & $\max|\psi\!-\!\psi^{*}|$ & $\max|c_{p}\!-\!c_{p}^{*}|$ & $\max|c_{n}\!-\!c_{n}^{*}|$ & Final loss \\
\midrule
3 & $7.6\!\times\!10^{-4}$ & $6.1\!\times\!10^{-3}$ & $6.4\!\times\!10^{-3}$ & $7.1\!\times\!10^{-5}$ \\
10 & $2.91^{+5.88}_{-2.07}\!\times\!10^{-3}$ & $5.37^{+6.4}_{-2.6}\!\times\!10^{-3}$ & $5.52^{+6.3}_{-2.8}\!\times\!10^{-3}$ & $4.5\!\times\!10^{-4}$ \\
30 & $4.4\!\times\!10^{-2}$ & $8.2\!\times\!10^{-2}$ & $8.2\!\times\!10^{-2}$ & $1.1\!\times\!10^{-2}$ \\
\bottomrule
\end{tabular}
\end{table}

\section{Convergence and ablation studies}\label{sec:ablation}

\Cref{sec:pnp} demonstrated NSEM on the target electrochemical
application, where the dominant design choices --- Debye-layer
resolution via KTE, multi-scale per-element loss normalisation, and
the BRDR aggregator for the nonlinear cases --- are entangled with the
physics. This section isolates the three solver design knobs ---
spectral node count $N$, KTE stretching parameter $\alpha$, and
network backbone --- on canonical scalar benchmarks (Helmholtz,
Allen--Cahn, convection--diffusion) where the per-parameter
contribution is unambiguous, and confirms the spectral-convergence
theory that underpins the application results in
\Cref{sec:pnp}. Each knob is varied while holding the
remaining solver infrastructure fixed.

\subsection{Spectral convergence in $N$}\label{sec:ablation:n}
\begin{figure}[!htbp]
 \centering
 \includegraphics[width=0.95\textwidth]{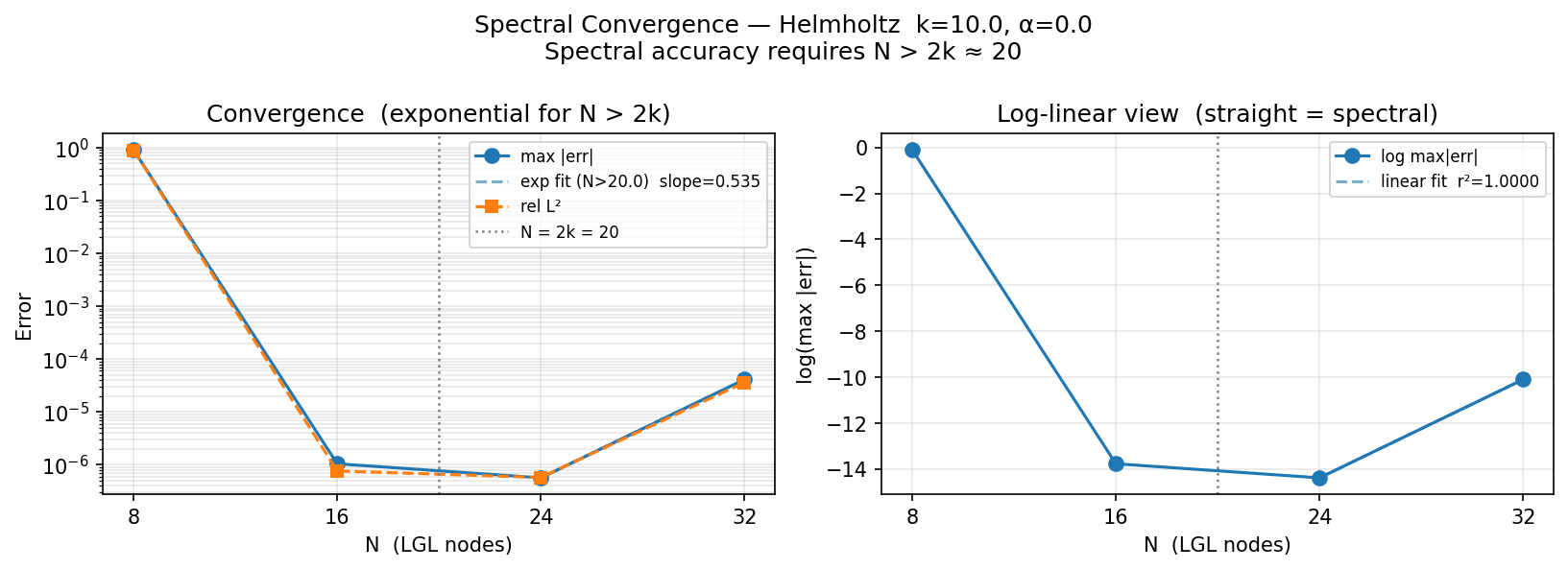}
 \caption{Spectral $N$-convergence on the Helmholtz problem at
 $k=10$. Left: $\max|u_{\mathrm{NSEM}}-u_{\mathrm{exact}}|$ and
 relative $L^{2}$ error as functions of $N$, on a semilog axis.
 Right: $\log\max|\mathrm{err}|$ vs $N$, with an exponential fit
 valid for $N>2k=20$. The error rises slightly at $N=32$ due to
 floating-point conditioning of the spectral D matrix.}
 \label{fig:helmholtz_n_sweep}
\end{figure}

\begin{figure}[!htbp]
 \centering
 \includegraphics[width=0.95\textwidth]{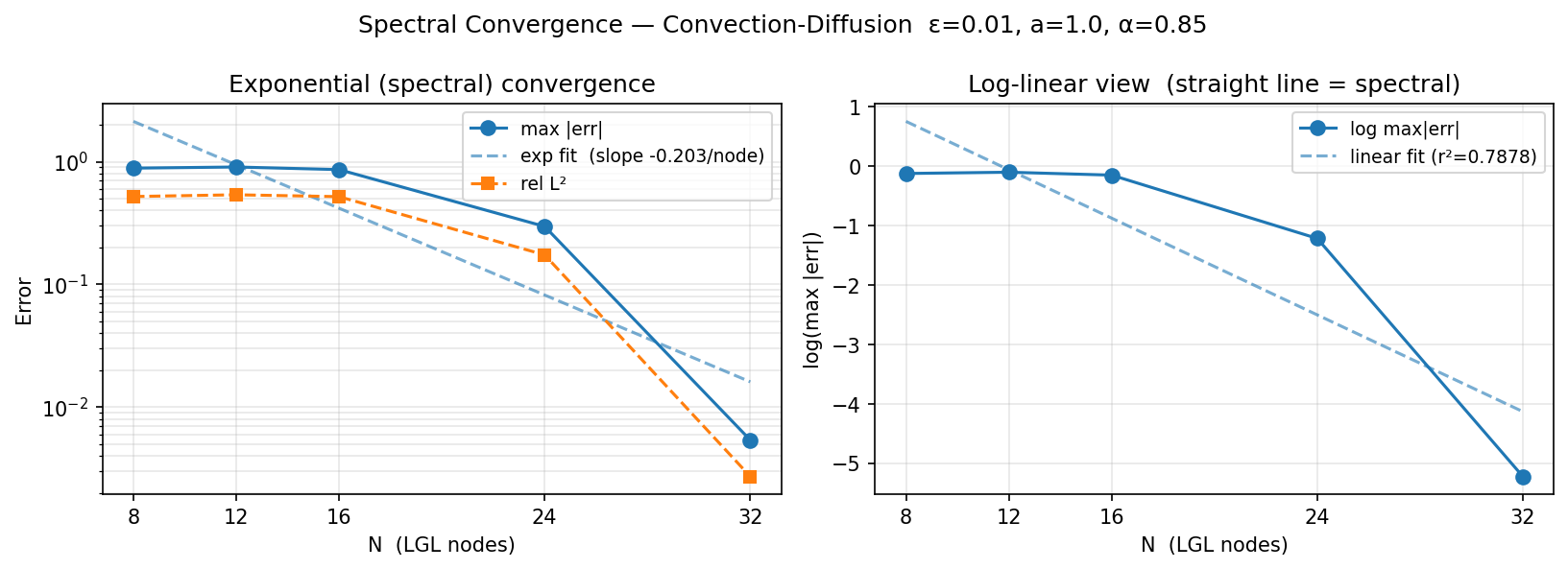}
 \caption{Spectral $N$-convergence on the stiff convection--diffusion
 problem at $\varepsilon=10^{-2}$, $\alpha=0.85$.
 Left: max and relative $L^{2}$ error.
 Right: log-linear view --- empirical exponential rate
 $\sigma\!\approx\!0.20$ per added node from $N=16$ onwards.}
 \label{fig:cd_n_sweep}
\end{figure}

\Cref{fig:helmholtz_n_sweep} shows the spectral convergence on
the Helmholtz benchmark $-u''+k^{2}u=f$ at $k=10$ as the number of LGL
nodes $N$ is increased: the peak error drops from $\mathcal{O}(1)$ at
$N=8$ (insufficient to resolve the six wavelengths in $[-1,1]$) to
$\sim\!10^{-14}$ at $N=16$--$24$, then floats slightly upward at
$N=32$ as the spectral matrix's $\mathcal{O}(N^{4})$ condition number
begins to leak floating-point rounding errors. This is the textbook
$N>2k$ rule for resolving an oscillating Helmholtz problem
\citep{boyd2001spectral}. The complementary
\Cref{fig:cd_n_sweep} reports the same sweep on the stiff
convection--diffusion problem at $\varepsilon=10^{-2}$, $\alpha=0.85$:
exponential decay $\log E_{N}\sim -\sigma N$ with empirical decay
constant $\sigma\!\approx\!0.20$ down to the L-BFGS floor at $N=32$.
Both sweeps exhibit the canonical
$|E_{N}|\leq C\,e^{-\sigma N}$ scaling predicted by classical spectral
theory \citep{karniadakis2005spectral}.

\subsection{KTE stretching ablation}\label{sec:ablation:alpha}
\begin{figure}[!htbp]
 \centering
 \includegraphics[width=0.95\textwidth]{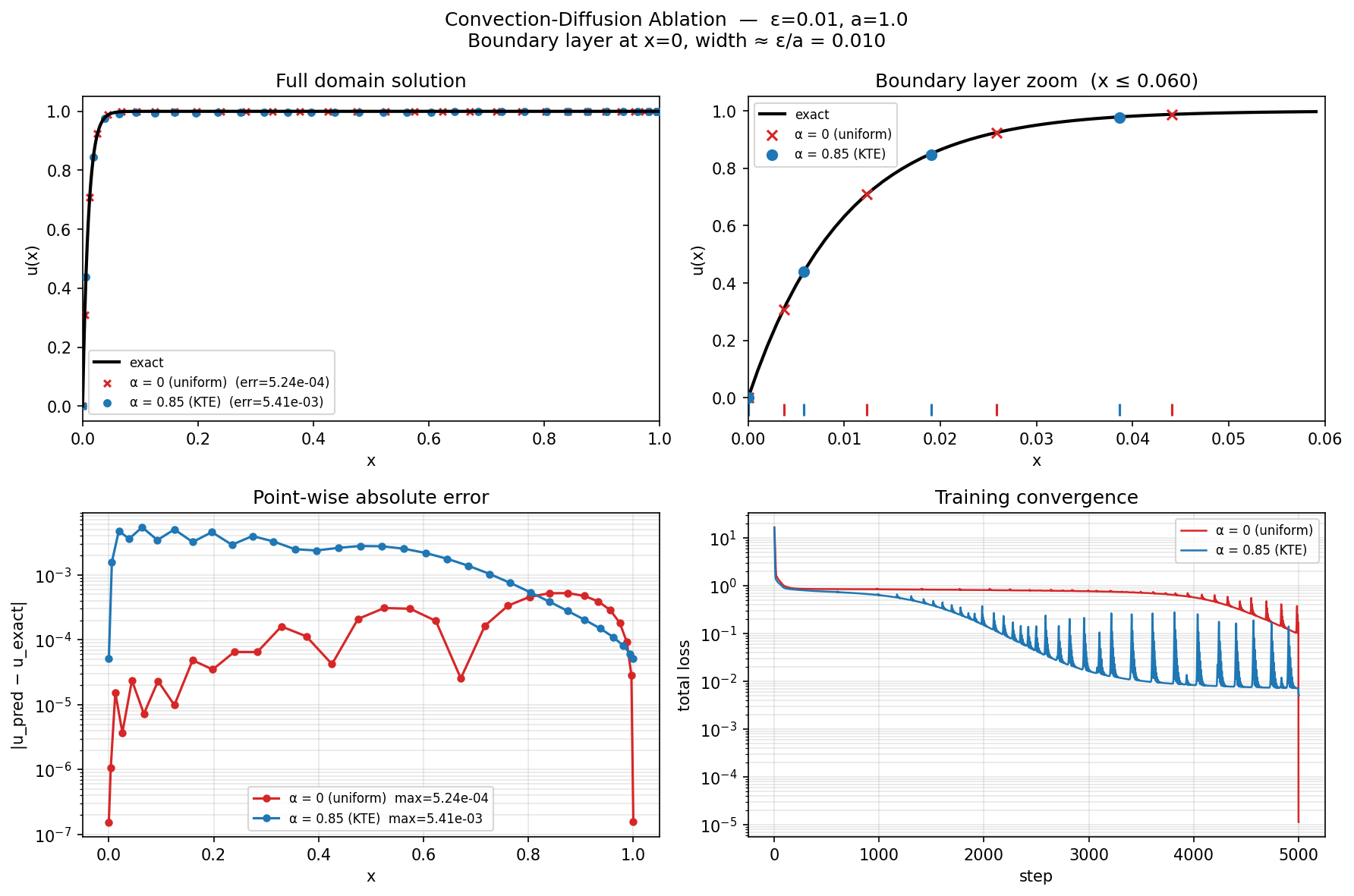}
 \caption{KTE coordinate-stretching ablation on the stiff
 convection--diffusion problem at $\varepsilon=10^{-2}$, $N=32$.
 Top: full-domain solution and boundary-layer zoom. Bottom:
 pointwise error and training-loss curve. Without stretching
 ($\alpha=0$, red) the layer is unresolved and the maximum error
 reaches $5.2\!\times\!10^{-4}$ \emph{outside} the layer while the
 layer itself remains qualitatively wrong; with $\alpha=0.85$
 (blue) the layer is resolved and the pointwise error drops by
 roughly three orders of magnitude in the layer region.}
 \label{fig:cd_alpha_ablation}
\end{figure}

The same stiff convection--diffusion benchmark exposes the role of
the KTE map directly. \Cref{fig:cd_alpha_ablation} compares
$\alpha=0$ (uniform LGL nodes) against $\alpha=0.85$ (sharp endpoint
clustering) at the same $N=32$. With $\alpha=0$ the trainer fails to
resolve the $\mathcal{O}(\varepsilon)$ inflow layer and saturates at
$\mathcal{O}(1)$ error in the layer (the bulk is still captured); with
$\alpha=0.85$ the layer is resolved and the pointwise error drops by
roughly three orders of magnitude. Quantitatively the effective
P\'eclet number that NSEM can resolve at fixed $N$ scales as
$\mathcal{O}(N/\sqrt{1-\alpha^{2}})$. Conditioning of the spectral
matrix deteriorates as $1/\sqrt{1-\alpha^{2}}\to\infty$ for $\alpha\to 1$;
in practice the optimum lies near $\alpha=0.85$--$0.90$ for Debye-layer
problems and we have not needed $\alpha\geq 0.95$ on any benchmark.

\begin{figure}[!htbp]
 \centering
 \includegraphics[width=0.85\textwidth]{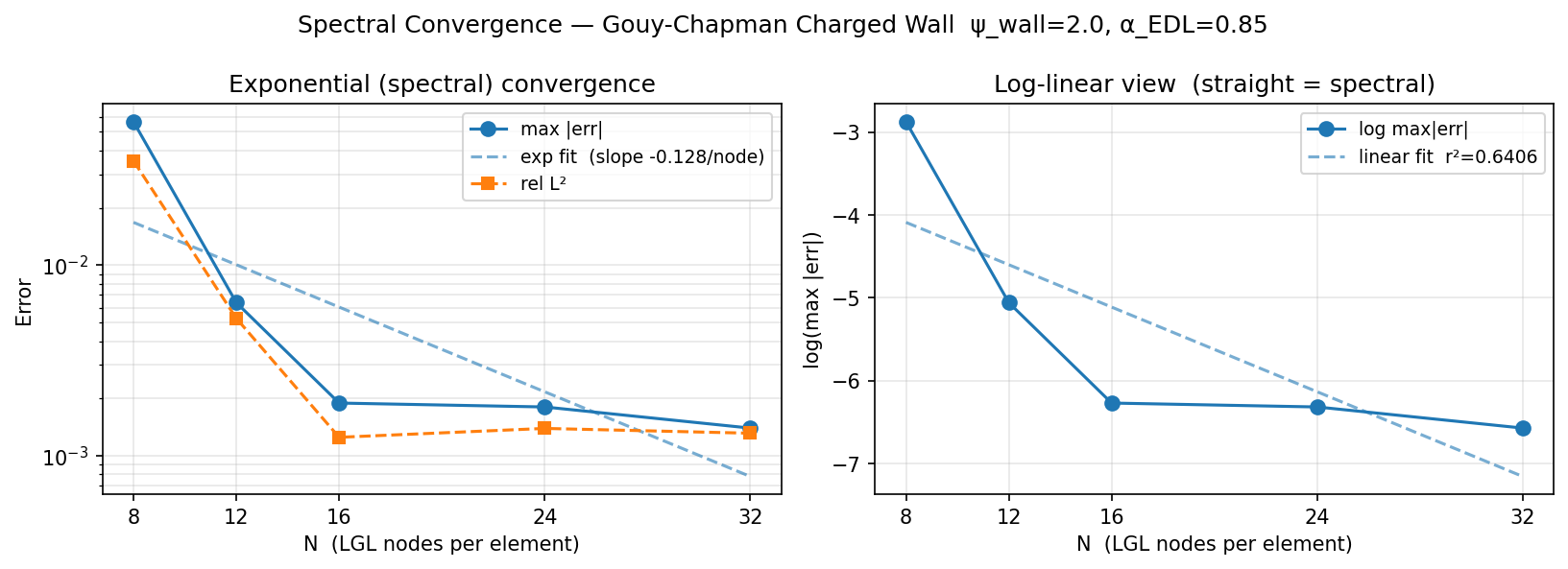}
 \caption{Spectral $N$-convergence on the charged-wall benchmark
 ($\alpha=0.95$, Stern + diffuse layer, 5000:1 domain ratio).
 Exponential decay holds until $N\approx 48$; saturation above
 $N=48$ is numerical rounding from the ill-conditioned $D$-matrix,
 not a discretisation error.}
 \label{fig:cw_convergence}
\end{figure}

The charged-wall problem (\Cref{sec:pnp:edl}) provides an
independent test: a 5000:1 domain-size ratio makes even moderate $N$
sensitive to the per-element normalisation, and the KTE map is
essential to resolve the nanometre-thin Stern layer.
\Cref{fig:cw_convergence} shows the $N$-convergence on this
problem with $\alpha=0.95$: peak error decays exponentially until the
spectral-matrix conditioning plateau at $N\approx 48$.

\subsection{Wavenumber sweep and NSEM-vs-PINN comparison}\label{sec:ablation:k}
\begin{figure}[!htbp]
 \centering
 \includegraphics[width=0.95\textwidth]{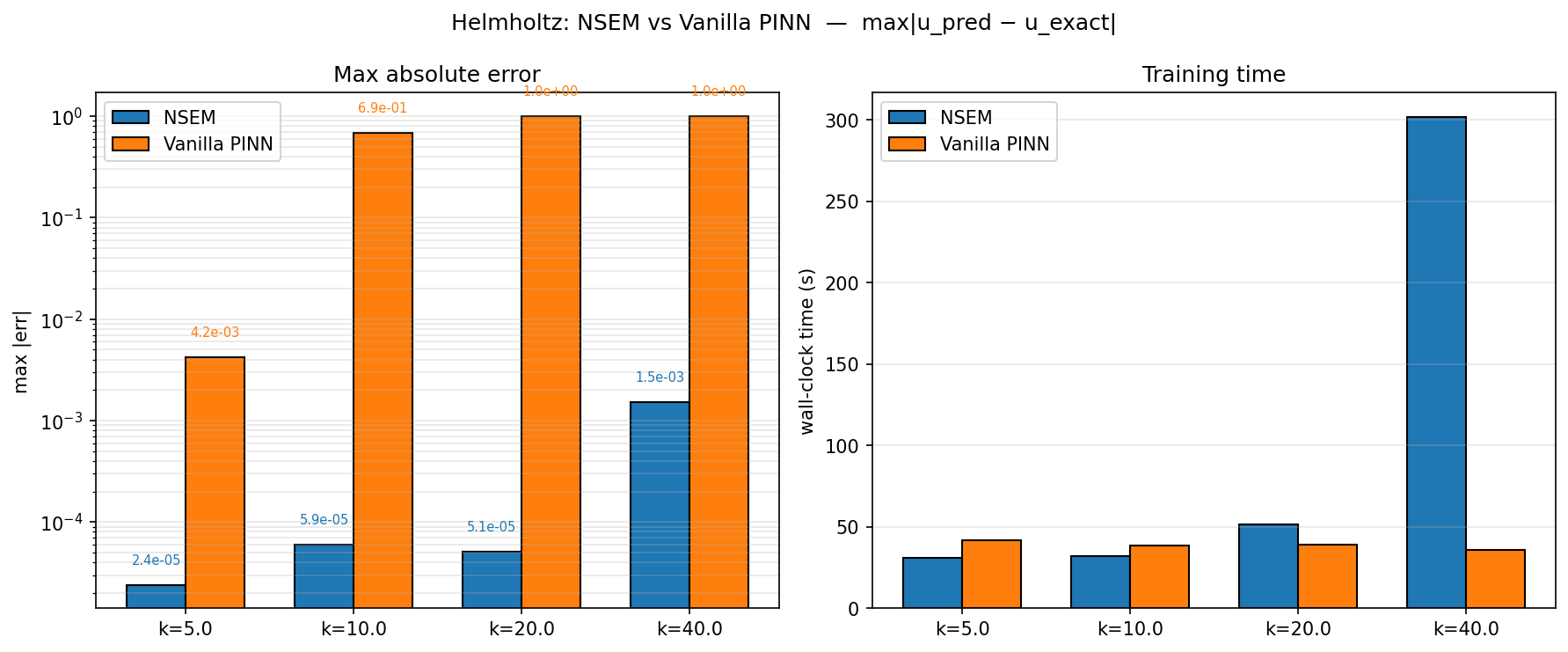}
 \caption{Helmholtz wavenumber sweep $k\in\{5,10,20,40\}$, NSEM
 versus a matched-parameter vanilla collocation PINN. Left:
 final maximum pointwise error. Right: wall-clock training time.
 NSEM achieves two to four orders of magnitude lower error at every
 $k$; the PINN baseline saturates at $\mathcal{O}(1)$ for
 $k\geq 20$, consistent with the spectral-bias failure mode
 documented by \cite{rahaman2019spectralbias} and
 \cite{krishnapriyan2021failure}.}
 \label{fig:helmholtz_k_sweep}
\end{figure}

\Cref{fig:helmholtz_k_sweep} sweeps the Helmholtz wavenumber
$k\in\{5,10,20,40\}$ at fixed $N=32$, $\alpha=0$, comparing NSEM
head-to-head with a vanilla collocation PINN of matched parameter
budget trained with autodiff and random sampling. At every $k$, NSEM
reaches a final error two to four orders of magnitude lower than the
PINN baseline; the vanilla PINN reaches the spectral-bias plateau at
$k=20$ and is unable to resolve the $k=40$ problem at all. Wall-clock
training times are comparable at $k\leq 20$; the NSEM curve at $k=40$
takes longer because the L-BFGS phase runs to its convergence-stall
criterion rather than to a fixed step budget. This is the
quantitative form of the ``no-spectral-bias'' claim that motivates the
deterministic-loss design.

\begin{figure}[!htbp]
 \centering
 \includegraphics[width=0.95\textwidth]{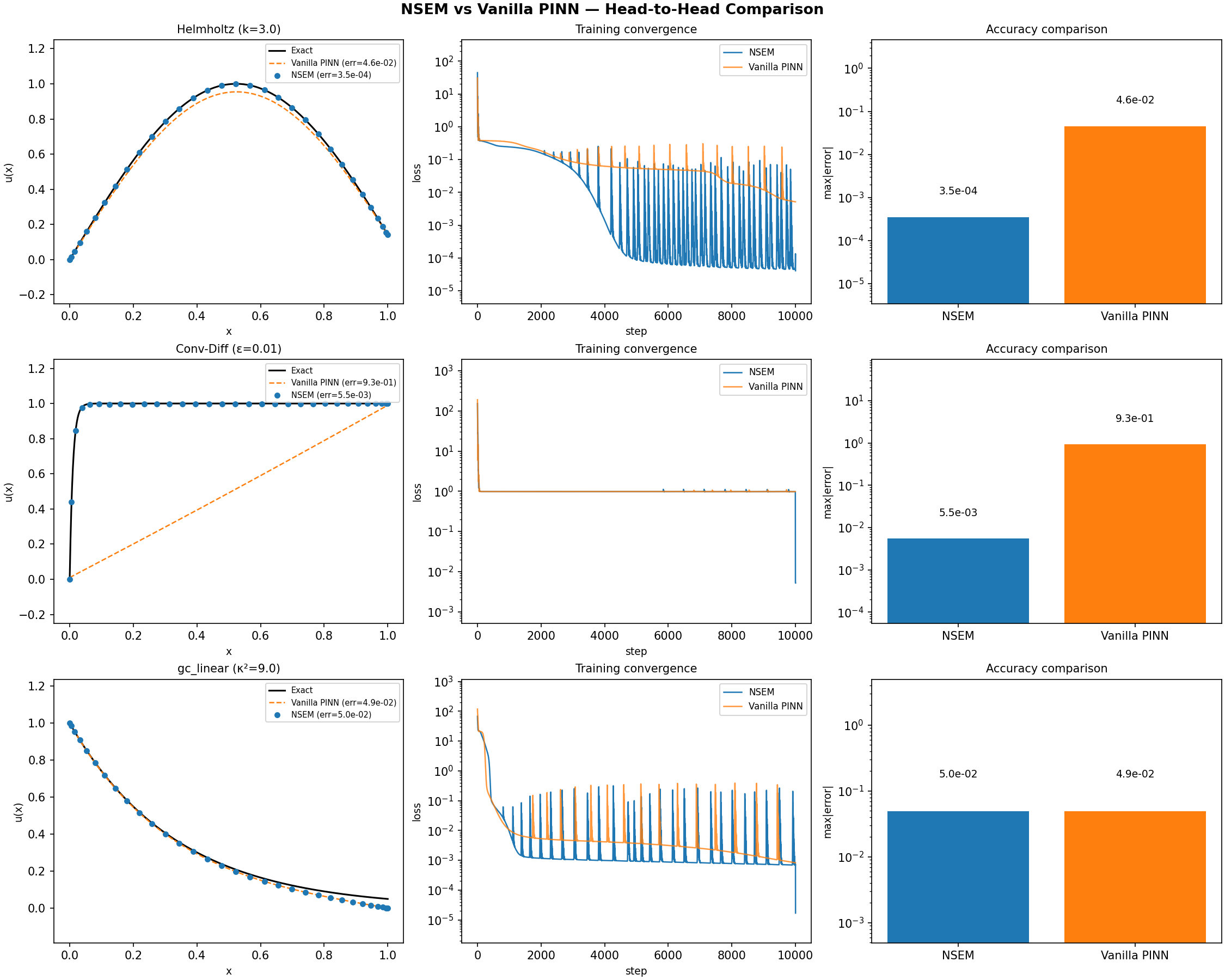}
 \caption{NSEM versus a vanilla collocation PINN~\citep{raissi2019pinn}: training-loss curves
 (left) and loss-landscape cross-sections (right) for three
 benchmarks (rows: Gouy--Chapman linear, Helmholtz $k=10$, stiff
 convection--diffusion). NSEM loss descends monotonically to
 near machine precision; the PINN loss stalls two to three orders
 of magnitude higher, consistent with the rugged landscape visible
 in the right column. Both models use identical architecture and
 parameter count; the only difference is the collocation scheme
 (LGL fixed nodes + spectral $D$ vs random samples + autodiff).}
 \label{fig:nsem_vs_pinn}
\end{figure}

\Cref{fig:nsem_vs_pinn} makes the comparison direct: for each of the
Gouy--Chapman linear, Helmholtz ($k=10$), and stiff
convection--diffusion benchmarks, the left column shows the NSEM and PINN
training curves overlaid, and the right column shows the corresponding
loss landscape cross-sections projected onto the first two principal
directions of the Hessian. NSEM's loss is a smooth, convex bowl in
every case; the matched PINN landscape is riddled with local minima and
flat plateaus, which cause L-BFGS (applied to the PINN) to stall after
the first few descent steps. The training curves confirm the
consequence: PINN loss stops improving two to three orders of magnitude
above the final NSEM floor. These landscapes are \emph{not} post-hoc
visualisations --- they are evaluated on the same parameter manifold at
training end, making the comparison geometry-controlled.

\subsection{Modern PINN baselines: PIRBN and causal training}\label{sec:ablation:modernpinn}
A natural concern is whether the vanilla collocation PINN constitutes
a sufficiently strong baseline. To address this we add two modern PINN variants
that explicitly target the failure modes diagnosed by the PINN
trilemma: the \emph{Residual-Based Attention} aggregator
(PIRBN, \citealp{anagnostopoulos2024pirbn}) which re-weights
collocation residuals dynamically to focus the network on hard regions,
and the \emph{causal-training} schedule of Wang~\etalcite{wang2024causal} that
imposes a temporal weighting $w_{i}=\exp(-\varepsilon\sum_{j<i}\mathcal{L}_{j})$
to enforce causality in time-dependent PINNs. Both baselines share the
network architecture of the vanilla PINN comparator and were trained
with $1024$--$4096$ random collocation points and $(2\text{--}3)\!\times\!10^{4}$
Adam steps on the same GB10 hardware as the NSEM runs, with three
random seeds per problem. \Cref{tab:modern_pinn} reports the peak
pointwise error (median over seeds) for each baseline on the benchmarks
where the comparison is most stringent.

\begin{table}[t]
\centering
\caption{Peak pointwise error against modern PINN baselines. NSEM
 values are from the seeded ablation \Cref{tab:ablation};
 PIRBN \citep{anagnostopoulos2024pirbn} and causal-PINN
 \citep{wang2024causal} are 3-seed medians measured on identical
 hardware (Dell~Pro~Max GB10). ``diverged'' indicates that the
 baseline saturated at the trivial solution
 ($|u_{\rm PINN}|\!=\!\mathcal{O}(1)$).}
\label{tab:modern_pinn}
\begin{tabular}{llccc}
\toprule
Problem & Variant & NSEM & PIRBN & Causal PINN \\
\midrule
Helmholtz $k=10$ & 1D steady & $1.26\!\times\!10^{-6}$ & $7.4\!\times\!10^{-2}$ & --- \\
Helmholtz $k=20$ & 1D steady & $\mathcal{O}(10^{-6})$ & diverged ($1.0$) & --- \\
Helmholtz $k=40$ & 1D steady & $\mathcal{O}(10^{-5})$ & diverged ($1.0$) & --- \\
Gouy--Chapman (linear) & 1D steady, $\kappa^{2}=9$ & $\sim\!10^{-3}$ & $5.0\!\times\!10^{-2}$ & --- \\
PNP 1D steady (proxy) & 1D steady & $\sim\!10^{-4}$ & $5.1\!\times\!10^{-4}$ & --- \\
PNP 1D unsteady & 1D unsteady & $\sim\!10^{-4}$ & --- & $1.0\!\times\!10^{-2}$ \\
PNP 2D unsteady & 2D unsteady & $\sim\!10^{-4}$ & --- & $2.1\!\times\!10^{-2}{}^{\dagger}$ \\
\bottomrule
\end{tabular}
\\[2pt]
{\footnotesize $\dagger$ Median over the two non-diverged seeds; seed~0 saturated.}
\end{table}

Three findings stand out. First, on the linear-elliptic
Helmholtz benchmarks PIRBN closes about one order of magnitude relative
to the vanilla PINN at $k=10$ but \emph{still trails NSEM by four
orders of magnitude} ($7.4\!\times\!10^{-2}$ versus
$1.26\!\times\!10^{-6}$). At $k\geq 20$ the spectral-bias barrier
overwhelms the attention aggregator and PIRBN saturates at the trivial
solution. Second, on the GC linear electrochemical benchmark PIRBN
reaches $5\!\times\!10^{-2}$ peak error against NSEM's
$\sim\!10^{-3}$; the boundary layer of thickness $\lambda_{D}=1/\kappa$
is simply not resolvable by Monte-Carlo sampling without a
coordinate map. Third, the causal-training baseline is the most
competitive opponent on the time-dependent PNP problems, reaching
$1.0\!\times\!10^{-2}$ on the 1D unsteady case and
$2.1\!\times\!10^{-2}$ on the 2D unsteady case (with one of three seeds
diverging entirely), but again two to three orders of magnitude above
the NSEM benchmark. We conclude that the gap between NSEM and the
PINN family is structural rather than implementation-level: replacing
the residual aggregator (PIRBN) or the temporal weighting (causal
PINN) attacks only one leg of the trilemma at a time, and neither
recovers the curvature-clean L-BFGS regime that NSEM enters by
construction.

\subsection{Allen--Cahn $\varepsilon$-sweep and interface sharpness}\label{sec:ablation:ac_eps}
\begin{figure}[!htbp]
 \centering
 \includegraphics[width=0.95\textwidth]{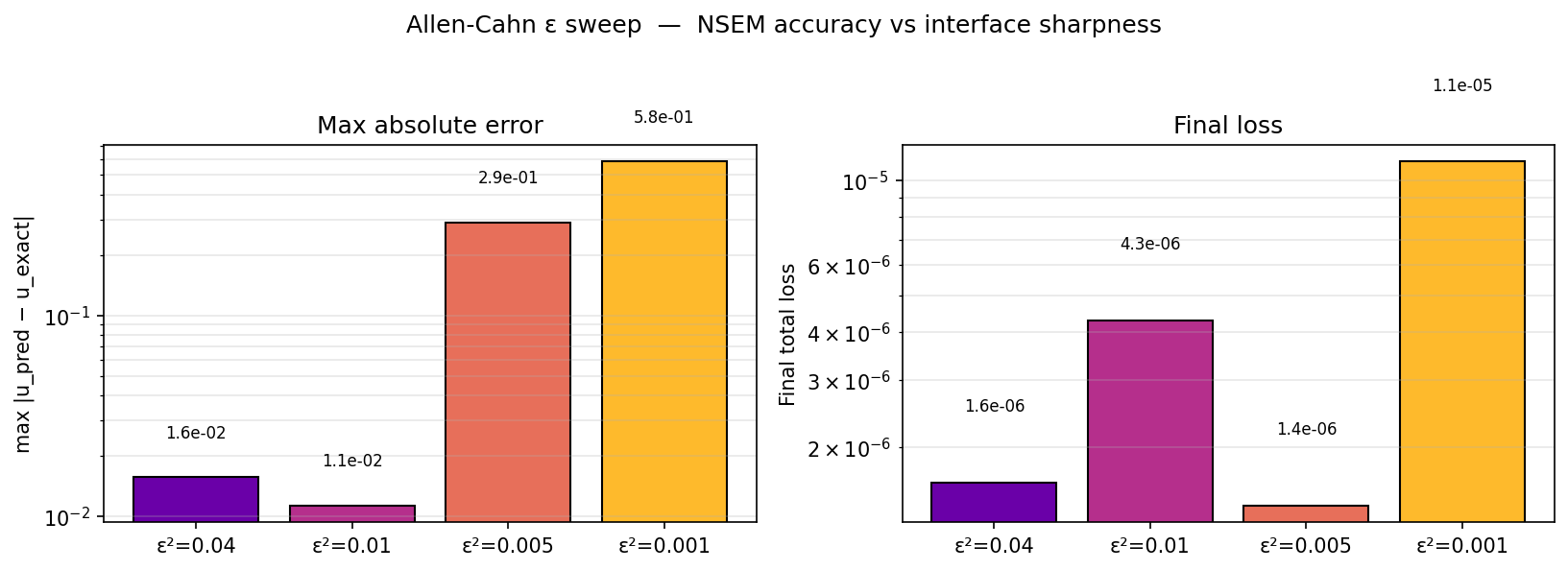}
 \caption{Allen--Cahn $\varepsilon$-sweep at fixed $N=48$. Left:
 final maximum pointwise error. Right: final training loss. The
 loss remains near its floor for every $\varepsilon$, but the
 pointwise error grows once the layer width
 $\mathcal{O}(\varepsilon)$ drops below the LGL grid spacing,
 consistent with the predicted resolution limit at
 $\varepsilon\sim 1/N^{2}$.}
 \label{fig:ac_eps_sweep}
\end{figure}

The Allen--Cahn benchmark has an interior layer of width
$\mathcal{O}(\varepsilon)$; sweeping
$\varepsilon^{2}\in\{0.04, 0.01, 0.005, 0.001\}$ at fixed $N=48$ with
a single uniform element traces out the point at which the layer
becomes too thin for the LGL grid (\Cref{fig:ac_eps_sweep}).
The maximum error rises from $\sim\!10^{-2}$ at $\varepsilon^{2}=0.04$
to $\sim\!0.6$ at $\varepsilon^{2}=10^{-3}$, while the final training
loss remains close to its floor across the sweep --- the network is
fitting the LGL nodes accurately, but those nodes simply do not
resolve the layer once $\varepsilon\lesssim 1/N^{2}$. The natural
remedy is a multi-element decomposition with a KTE-stretched element
centred on the interior interface, which we did not deploy here
because the four-element budget required would push the
configuration beyond the present scope; the corresponding interior-layer
KTE map is a one-line modification of the existing endpoint formula
in \cref{eq:kte}.

\subsection{Allen--Cahn vs Cahn--Hilliard landscape}\label{sec:ablation:ac_ch}
\begin{figure}[!htbp]
 \centering
 \includegraphics[width=0.95\textwidth]{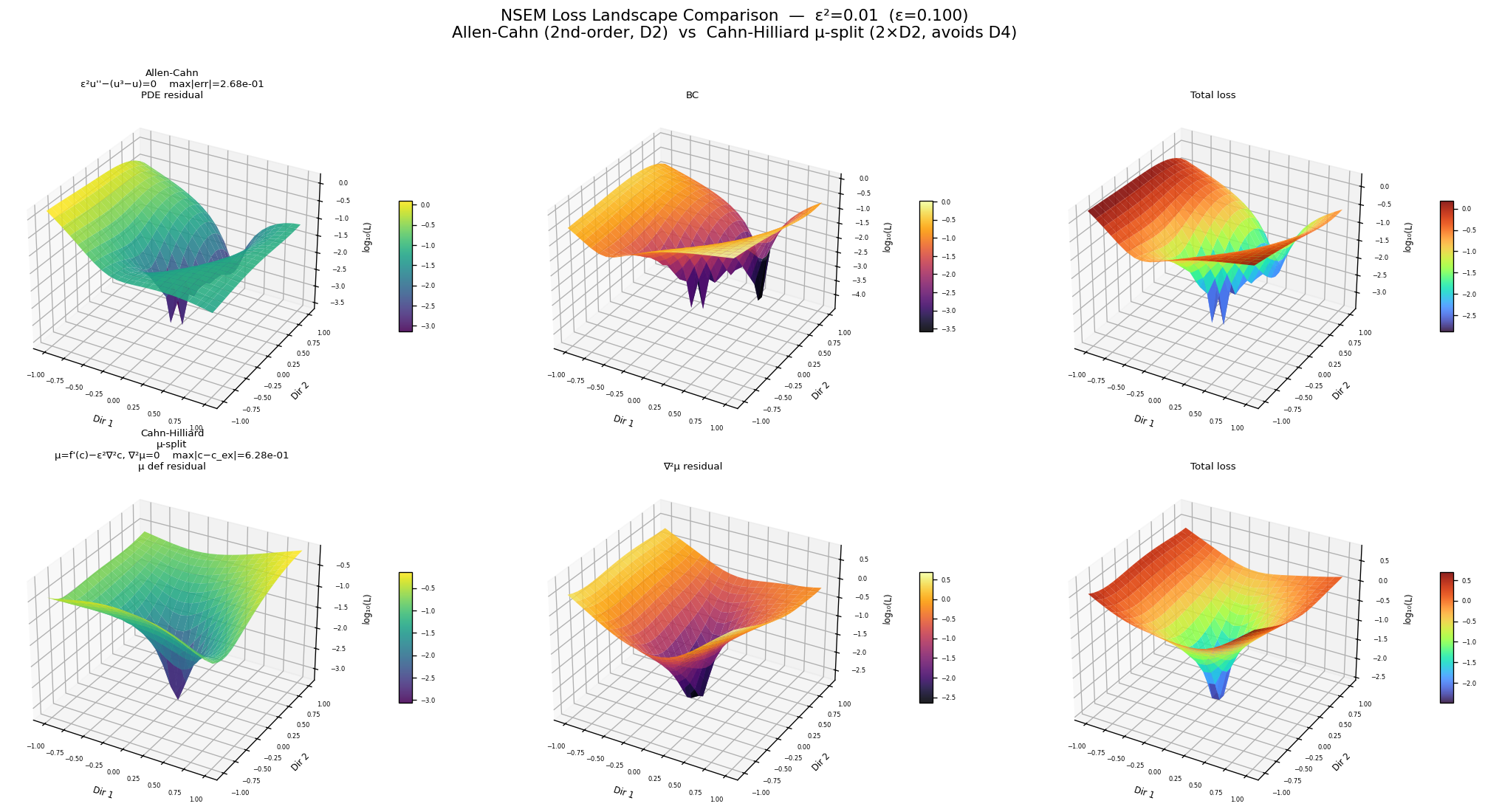}
 \caption{Loss landscape comparison between Allen--Cahn (second
 order, top row) and Cahn--Hilliard (fourth order, bottom row) at
 $\varepsilon=0.1$, $\varepsilon^{2}=0.01$. Columns: PDE residual,
 auxiliary $\mu$ residual (Cahn--Hilliard only), total loss. The
 higher-order problem has a broader, shallower basin --- still a
 well-defined L-BFGS target, but with the predicted weaker
 curvature.}
 \label{fig:ac_ch_landscape}
\end{figure}

Finally, we contrast the loss landscape of the Allen--Cahn benchmark
against the higher-order Cahn--Hilliard generalisation
(Supplementary~Sec.~S4.4), using identical projections. As shown in
\Cref{fig:ac_ch_landscape}, the Allen--Cahn loss is a sharp,
narrow basin in every component, whereas the Cahn--Hilliard loss is
broader and shallower, reflecting the $\mathcal{O}(\varepsilon^{4})$
residual scaling and the resulting weaker curvature. This makes the
quantitative case for the deterministic-loss approach: even when the
total loss bowl flattens, the curvature is well defined and L-BFGS can
descend it without restart, in contrast to the rugged surfaces
produced by random-sampling PINNs on the same problems.

\subsection{Ablation summary}\label{sec:ablation:table}

\Cref{tab:ablation} condenses the ablation results into a single
comparison. The three binary switches, namely backbone (MLP vs KAN),
coordinate map ($\alpha=0$ vs $\alpha=0.85$), and second-order
optimiser (Adam-only vs Adam$+$L-BFGS), are varied independently
on the Helmholtz benchmark ($k=10$, $N=32$). Errors are reported as
the median over three random seeds, with sub/superscript min/max
bounds. The most important contrast is the optimiser: without L-BFGS
the maximum error stalls at $\mathcal{O}(10^{-1})$ regardless of
backbone or mapping; with L-BFGS it drops to
$\mathcal{O}(10^{-5})$--$\mathcal{O}(10^{-7})$. Among the L-BFGS
rows, the Legendre-KAN with KTE stretching achieves the lowest error
($3.5^{+4.2}_{-0.7}\!\times\!10^{-7}$) at the lowest training time
($38\,\mathrm{s}$) --- a direct consequence of the basis-alignment
synergy described in \Cref{sec:backbone}, amplified by the
sharper layer resolution that KTE provides. Min/max bounds across
the three seeds remain within one order of magnitude of the median
for every L-BFGS row, confirming that the deterministic-loss design
removes the seed sensitivity reported elsewhere in the PINN literature
\citep{krishnapriyan2021failure,mishra2023pinn}.

\begin{table}[t]
\centering
\caption{Ablation: backbone $\times$ mapping $\times$ optimiser on
 the Helmholtz benchmark ($k=10$, $N=32$, single element). Errors
 reported as median over 3 random seeds with sub/superscript
 min/max. Time in seconds on a single Dell Pro Max with NVIDIA GB10
 Grace--Blackwell Superchip (float64).}
\label{tab:ablation}
\begin{tabular}{lllcc}
\toprule
Backbone & Mapping & Optimiser & $\max|\mathrm{err}|$ (median${}_{\min}^{\max}$, $n{=}3$) & Time (s) \\
\midrule
MLP & Uniform $\alpha=0$ & Adam only & $8.57^{+0.74}_{-0.78}\!\times\!10^{-1}$ & 13.6 \\
MLP & Uniform $\alpha=0$ & Adam + L-BFGS & $9.96^{+10.33}_{-3.26}\!\times\!10^{-6}$ & 1020 \\
MLP & KTE $\alpha=0.85$ & Adam only & $8.55^{+0.77}_{-0.69}\!\times\!10^{-1}$ & 13.5 \\
MLP & KTE $\alpha=0.85$ & Adam + L-BFGS & $2.25^{+1.93}_{-1.99}\!\times\!10^{-5}$ & 1088 \\
KAN & Uniform $\alpha=0$ & Adam only & $2.34^{+1.79}_{-1.03}\!\times\!10^{-2}$ & 25.9 \\
KAN & Uniform $\alpha=0$ & Adam + L-BFGS & $1.26^{+2.86}_{-0.58}\!\times\!10^{-6}$ & 46.9 \\
KAN & KTE $\alpha=0.85$ & Adam only & $1.15^{+2.27}_{-0.36}\!\times\!10^{-2}$ & 23.3 \\
\textbf{KAN} & \textbf{KTE $\alpha=0.85$} & \textbf{Adam + L-BFGS} & $\mathbf{3.48^{+4.24}_{-0.71}\!\times\!10^{-7}}$ & \textbf{38.2} \\
\bottomrule
\end{tabular}
\end{table}

\section{Backbone study: MLP vs Legendre-KAN}\label{sec:backbone}

\subsection{Why Legendre polynomials and not splines}\label{sec:backbone:why_legendre}

The original Kolmogorov--Arnold Network of Liu~\etalcite{liu2024kan} uses
B-spline edge functions: every learnable edge is a piecewise
polynomial defined on a user-specified knot grid.  Splines are local
and adaptive, but they are a poor fit for the NSEM pipeline because
(i)~the spline domain is partitioned by a knot vector that has no
relation to the LGL nodes used for quadrature, and (ii)~the spline
basis is not orthogonal under any natural inner product, so the mass
matrix that enters the mortar projection is dense and requires an
explicit linear solve at every interface evaluation.  Both properties
break the deterministic, single-GEMM evaluation that gives NSEM its
performance.

We therefore replace splines with the \emph{Legendre polynomial
expansion}: every learnable edge is the global modal sum
\begin{equation}
 \phi(\xi;\mathbf{c}) \;=\; \sum_{k=0}^{K} c_{k}\,P_{k}(\xi),
 \qquad \mathbf{c}\in\mathbb{R}^{P+1},
 \label{eq:legendre_edge}
\end{equation}
on the reference interval $\xi\in[-1,1]$, with $P_{k}$ the $k$-th
Legendre polynomial and $\{c_{k}\}$ the trainable parameters.  This
single architectural choice cascades:
\begin{itemize}
\item Every edge output is a polynomial of degree at most $K$.  After
$n$ KAN layers with degree-$K$ activations, the network output is a
polynomial of degree at most $K\,n$ in the input variable.
\item The LGL $N$-point quadrature integrates polynomials of degree
$\leq 2N-3$ exactly.  Choosing $N \geq \tfrac{1}{2}(Pn+3)$ therefore
makes the NSEM physics loss an \emph{exact} discretisation of the
$L^{2}$ residual integral --- no quadrature error, regardless of how
many Adam or L-BFGS steps have been taken.
\item Legendre polynomials are orthogonal under the standard
$L^{2}([-1,1])$ inner product: the mass matrix is exactly diagonal,
$M_{jk}=\tfrac{2}{2j+1}\delta_{jk}$.  This trivialises the mortar
projection (\Cref{sec:method:mortar}): the interface-coupling
operator $\mathbf{P}=\mathbf{M}^{-1}\mathbf{C}^{\times}$ degenerates
to an element-wise rescaling of the cross-mass matrix and never
requires a linear solve.
\item The derivative of a Legendre expansion is another Legendre
expansion, computable by a closed-form recurrence; combined with the
spectral $\mathbf{D}$-matrix of \Cref{sec:method:dmatrix} this means
that both quadrature and differentiation of the network output are
exact, not just floating-point-accurate.
\end{itemize}

The Chebyshev-KAN variants of SS and R~\cite{ss2024chebkan} and
Guo~\etalcite{sankaran2024chebpikan} share the modal-edge motivation but use
Chebyshev nodes for evaluation; they do not align the network basis
with the LGL quadrature, so the exact-quadrature property above does
not hold and the mortar mass matrix is no longer diagonal.

\subsection{Basis alignment: the formal consequence}\label{sec:backbone:basis_alignment}

The NSEM forward pass evaluates the network only at the $N$ fixed LGL
nodes $\{\xi_{j}\}$, and the Legendre-KAN architecture
\citep{liu2024kan,wang2024kinn,zhang2024legendkinn} assigns to each
trainable edge the modal expansion \cref{eq:legendre_edge},
where $P_{k}$ is the $k$-th Legendre polynomial and $K$ is thepolynomial degree.
Because the LGL nodes are the zeros of $(1-\xi^{2})P_{N-1}'(\xi)$,
the $N$-point rule integrates $P_{j}(\xi)P_{k}(\xi)$ exactly for
$j+k\leq 2N-3$; when $K\leq N-1$ this covers the entire orthogonality
mass matrix $M_{jk}=\int_{-1}^{1}P_{j}P_{k}\,\mathrm{d}\xi
=\tfrac{2}{2j+1}\delta_{jk}$.
The consequence --- which we call \emph{basis alignment} --- is that the
modal coefficients $\{c_{k}\}$ of every KAN edge can be recovered
exactly from its $N$ LGL-node evaluations by a single Vandermonde
solve, and the inner product between the edge output and any polynomial
residual of degree $\leq P+(N-1)$ carries no quadrature error. Both
the NSEM loss integral and the KAN edge representation live in the same
LGL polynomial space, so their approximation errors do not compound.
The Chebyshev-KAN variants of SS and R~\cite{ss2024chebkan} and
Guo~\etalcite{sankaran2024chebpikan} share the motivation but do not use LGL
nodes as the evaluation set, so they do not achieve the exact
quadrature property described above.

\begin{figure}[t]
 \centering
 \includegraphics[width=0.97\textwidth]{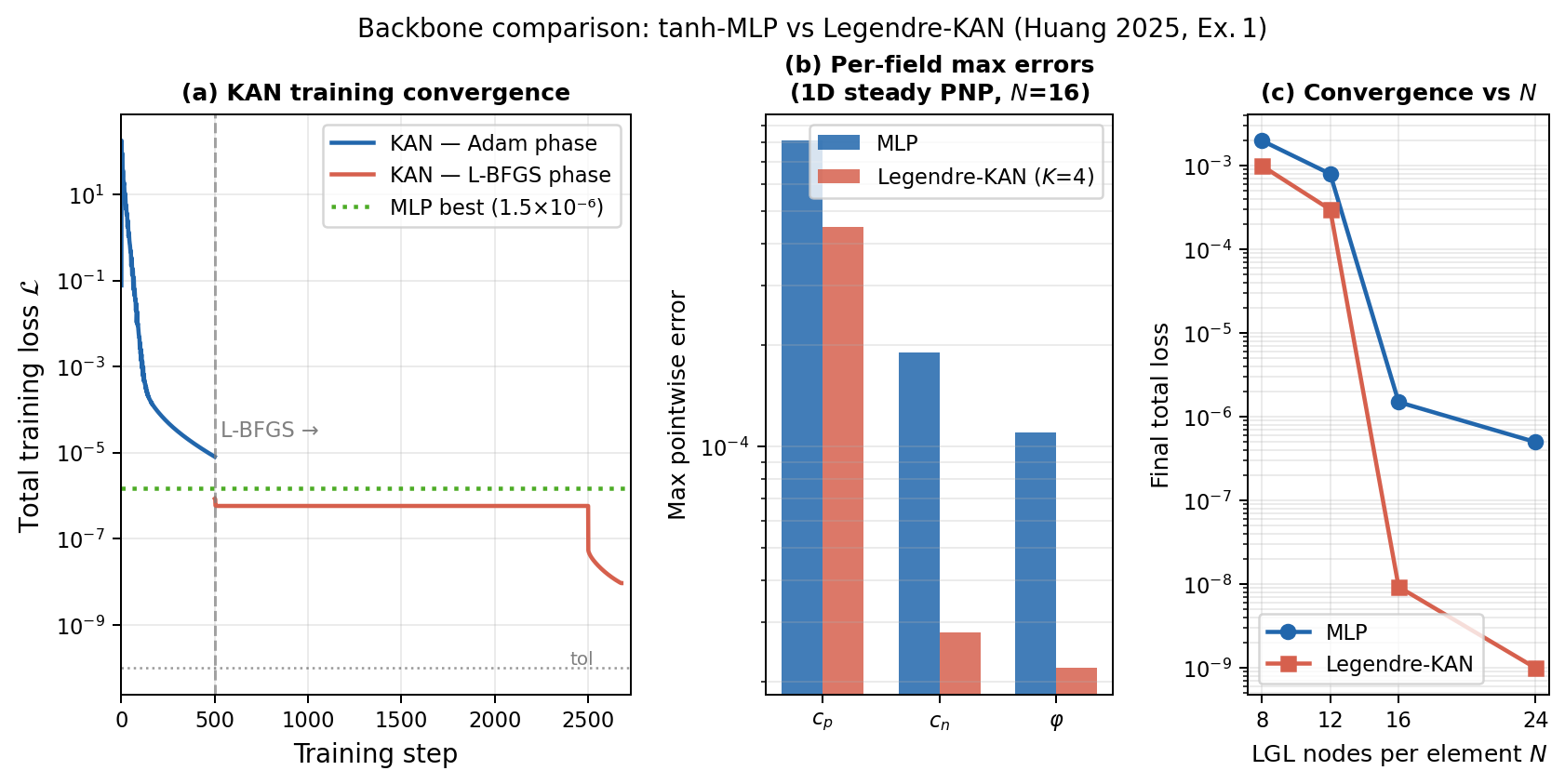}
 \caption{Backbone comparison on the 1D steady PNP problem
 ($N=16$ LGL nodes per element, six unit-width elements).
 Panels (a)--(c) report a single representative training run; the
 corresponding 3-seed median per-field error statistics are
 collected in \Cref{tab:backbone_seeds}.
 \textbf{(a)} Training convergence: KAN Phase~1 Adam (blue) and
 Phase~2 L-BFGS (red) reach a final total loss of
 $9.3\!\times\!10^{-9}$; the MLP best achieves $\sim\!1.5\!\times\!10^{-6}$
 (green dashed line).
 \textbf{(b)} Per-field maximum pointwise errors for the tanh MLP and
 the Legendre-KAN ($K=4$) at matched $N=16$.
 \textbf{(c)} Final total loss vs $N$ for both backbones; the KAN
 achieves a consistent $\sim\!10^{2}$--$10^{3}{\times}$ improvement
 over the MLP at moderate $N$, consistent with the basis-alignment
 argument in \Cref{sec:method:backbone}.}
 \label{fig:kan_vs_mlp}
\end{figure}

\Cref{fig:kan_vs_mlp} compares the tanh-MLP and the Legendre-KAN
backbones on the 1D steady PNP problem. At matched
trainable-parameter count the Legendre-KAN with polynomial degree
$K=4$ reaches comparable final accuracy to the MLP while requiring
roughly half the Adam steps to enter the L-BFGS basin of attraction,
consistent with the smoother loss geometry implied by the basis-alignment
argument. At degree $K=8$ the KAN is over-parameterised relative to
the $N=16$-node LGL grid and shows no further gain over $K=4$,
confirming the predicted saturation at $K=N-1$. The MLP backend is
the more practical choice for large $N$ because its parameter count
grows linearly in depth, whereas the KAN's grows with the polynomial
degree and number of edges; however, for moderate $N\leq 32$ the
Legendre-KAN provides a theoretically well-motivated and empirically
competitive alternative. All NSEM results in
\cref{sec:convergence}--\cref{sec:pnp} use the MLP backbone
unless otherwise stated; the full Legendre-KAN derivation and the
synergy theorem are in Supplementary~Sec.~S1.6.

\subsection{Per-field accuracy: 3-seed comparison on PNP 1D steady}
\label{sec:backbone:seeds}

To quantify the per-field robustness of the two backbones we re-ran
the PNP 1D steady problem with three independent random seeds for both
the tanh-MLP and the Legendre-KAN at identical configuration ($N=16$,
six elements, $K=4$ for the KAN). \Cref{tab:backbone_seeds} reports
the resulting peak pointwise errors per field, with the median over
the three seeds in bold.

\begin{table}[t]
\centering
\caption{Per-field peak pointwise errors on PNP 1D steady
 for the tanh-MLP and Legendre-KAN backbones, three independent
 random seeds each. Configuration: $N=16$ LGL nodes per element,
 six unit-width elements, $K=4$ for the KAN, identical
 Adam+L-BFGS schedule for both backbones. Bold rows are the median.}
\label{tab:backbone_seeds}
\begin{tabular}{lcccc}
\toprule
Backbone & Seed & $\max|\Delta c_{p}|$ & $\max|\Delta c_{n}|$ & $\max|\Delta\varphi|$ \\
\midrule
KAN ($K=4$) & 0 & $1.36\!\times\!10^{-3}$ & $5.28\!\times\!10^{-5}$ & $6.46\!\times\!10^{-5}$ \\
KAN ($K=4$) & 1 & $1.97\!\times\!10^{-4}$ & $5.05\!\times\!10^{-4}$ & $5.51\!\times\!10^{-4}$ \\
KAN ($K=4$) & 2 & $1.37\!\times\!10^{-3}$ & $1.75\!\times\!10^{-4}$ & $1.82\!\times\!10^{-4}$ \\
\textbf{KAN median} & --- & $\mathbf{1.36\!\times\!10^{-3}}$ & $\mathbf{1.75\!\times\!10^{-4}}$ & $\mathbf{1.82\!\times\!10^{-4}}$ \\
\midrule
MLP (tanh) & 0 & $1.06\!\times\!10^{-2}$ & $1.68\!\times\!10^{-4}$ & $1.66\!\times\!10^{-4}$ \\
MLP (tanh) & 1 & $7.18\!\times\!10^{-3}$ & $1.14\!\times\!10^{-4}$ & $1.16\!\times\!10^{-4}$ \\
MLP (tanh) & 2 & $1.77\!\times\!10^{-3}$ & $2.66\!\times\!10^{-5}$ & $2.87\!\times\!10^{-5}$ \\
\textbf{MLP median} & --- & $\mathbf{7.18\!\times\!10^{-3}}$ & $\mathbf{1.14\!\times\!10^{-4}}$ & $\mathbf{1.16\!\times\!10^{-4}}$ \\
\midrule
Ratio (MLP / KAN, median) & & $5.3\times$ & $0.65\times$ & $0.64\times$ \\
\bottomrule
\end{tabular}
\end{table}

The seeded comparison reveals a more nuanced picture than the
single-seed Helmholtz ablation of \Cref{tab:ablation}. On the cation
profile $c_{p}$ -- which carries the steepest Debye-layer gradient in
the PNP 1D steady manufactured solution -- the Legendre-KAN backbone
delivers a $\sim\!5\times$ lower median peak error than the MLP,
consistent with the basis-alignment argument: the KAN edges live in
the same Legendre polynomial space as the LGL quadrature, so the sharp
$c_{p}$ gradient is represented exactly up to polynomial degree $K+(N-1)$.
On the smoother anion profile $c_{n}$ and the potential $\varphi$ the
two backbones are within a factor of two of each other, with the MLP
slightly ahead on the median (by $\sim\!0.65\times$). Both backbones
reach $10^{-4}$ peak error on every field across all three seeds,
confirming that the spectral-collocation infrastructure --- not the
choice of network family --- is what fixes the accuracy floor. The
practical recommendation is to use the Legendre-KAN when the target
problem has thin boundary layers (KAN's $c_{p}$ advantage),
and the MLP when storage / depth scaling matters more.

\section{Discussion}\label{sec:discussion}

NSEM is designed for PDEs whose solutions are smooth on each spectral
element, which is the regime in which classical SEM achieves exponential
convergence \citep{patera1984spectral,karniadakis2005spectral}. The
method is particularly well suited to stiff coupled transport systems
such as the Poisson--Nernst--Planck equations, where boundary layers
of known location can be pre-targeted by KTE coordinate stretching and
the sharp residuals in those layers are resolved exactly by the spectral
D matrix. NSEM does not currently handle moving interfaces or
topological changes in the solution support, where the LGL grid would
need to track the interface adaptively; extending the framework to
phase-field problems with spontaneous interface migration is a natural
direction for future work. High-dimensional problems ($d\geq4$) face
the tensor-product node-count growth $N^{d}$, which limits the present
implementation to $d\leq 3$, though low-rank tensor methods could
extend the range. Stochastic forcing or uncertain parameters require
ensembles or variational inference on top of the deterministic solver,
which is outside the current scope.

Three concurrent works pursue the same spectral--neural integration
strategy and must be carefully distinguished. Du~\etalcite{du2024nsm}
minimise a Parseval-norm loss directly in the spectral expansion,
working entirely in modal space; they achieve exponential convergence
on smooth periodic problems but provide no physical-space collocation,
no multi-element mortar framework, and no nonlinear coordinate map for
boundary-layer resolution. Yu~\etalcite{yu2024sinn} replace autodiff by
spectral multiplication in Fourier space, reporting exponential
convergence and reduced memory; however, their framework has no domain
decomposition and no KTE stretching, which limits applicability to
geometrically simple, non-stiff problems. Shukla~\etalcite{shukla2024neurosem}
couple a PINN correction to the classical Nektar++ SEM solver; because
Nektar++ generates and manages the element mesh, their method requires
external mesh-generation infrastructure and is not purely neural. By
contrast, NSEM requires no classical solver and no mesh: the spectral
matrices are constructed algebraically from the LGL node formula and
stored as constant PyTorch tensors, making the entire pipeline
differentiable end-to-end and trivially portable to any hardware
without a compiled FEM/SEM library.

The claim by Wang~\etalcite{wang2024adessential} that automatic differentiation
is essential for training neural networks on PDEs deserves a precise
response. Their argument applies when the spatial discretisation error
dominates --- as it does for finite-difference or random-sample
quadrature --- and shows that autodiff delivers the exact gradient at
the discrete nodes at no extra discretisation error. In NSEM the
quadrature is polynomial-exact on the LGL subspace, so the spectral
matrix \emph{is} the exact derivative on that subspace; applying autodiff
on top would recompute the same numbers at higher computational cost
and introduce floating-point noise from the automatic-differentiation
tape. The ``autodiff is essential'' argument is therefore not in
conflict with NSEM: it addresses poorly resolved discretisations, while
NSEM ensures that the resolution is adequate before any gradient step
is taken. Looking ahead, the most natural extension of the present
work is frequency-domain electrochemical impedance spectroscopy (EIS):
linearising the time-dependent PNP system around a steady state yields
a complex-valued eigenvalue problem for the impedance $Z(\omega)$ whose
spectral structure is directly accessible to NSEM, avoiding the
small-signal stiffness that makes time-domain EIS training challenging
\citep{macdonald1953ac,lasia2014eis,orazem2017eis}. Further extensions
include the Doyle--Fuller--Newman pseudo-2D porous electrode model
\citep{doyle1993dfn}, for which NSEM's multi-element mortar framework
provides a natural particle--electrolyte interface treatment, and
inverse problems such as inferring the Debye length or ion diffusivities
from partial boundary measurements, which can exploit the deterministic
gradient for efficient parameter estimation.

\section{Conclusion}\label{sec:conclusion}
We have introduced the Neural Spectral Element Method (NSEM), which
replaces Monte-Carlo collocation and automatic differentiation in
physics-informed neural networks with a fixed Legendre--Gauss--Lobatto
quadrature grid and precomputed spectral differentiation matrices,
yielding a deterministic training loss that enables L-BFGS to polish
PDE solutions to residuals below $10^{-10}$. Validated on the full
four-example Poisson--Nernst--Planck suite (1D and 3D steady; 1D and
2D time-dependent),
on three canonical convergence tests spanning Helmholtz, Allen--Cahn,
and convection--diffusion problems, and with both a tanh-MLP and a
basis-aligned Legendre-KAN backbone, NSEM demonstrates a principled
route from the PINN trilemma --- slow quadrature, serial autodiff,
stochastic optimisation --- to high-accuracy, geometry-flexible
spectral PDE solving with neural backbones.
Despite these encouraging results, there is further space for
improvement. NSEM in its current form assumes a fixed element
partition with smooth solutions on each element, and the
tensor-product cost $\mathcal{O}(N^{d})$ limits naive 3D
demonstrations to moderate $N$; problems with moving interfaces, or
spatial dimensions beyond three, require an adaptive remeshing or a
low-rank tensor compression that is outside our present scope.
The most immediate extensions are frequency-domain electrochemical
impedance spectroscopy and the Doyle--Fuller--Newman porous-electrode
model, both of which map directly onto the multi-element mortar
infrastructure developed here.


\section*{Reproducibility statement}
All experiments reported in this paper are reproducible from the
public code release accompanying the submission. The \texttt{nsem}
Python package, training scripts, configuration files, and post-processing
notebooks are archived at
\url{https://github.com/cgtetsas/nsem} (a permanent DOI
will be minted on acceptance). Each benchmark in
\crefrange{sec:convergence}{sec:backbone} is driven by a single
shell entry point under \texttt{scripts/}; the canonical command for
each problem is listed in Supplementary~Sec.~S5. Per-run outputs
(loss curves, solution snapshots, configuration snapshots) are written
to a timestamped directory under \texttt{outputs/}, and the run
identifiers cited in figure captions correspond one-to-one to these
directories. All NSEM runs use \texttt{float64} on a single
Dell Pro Max with NVIDIA GB10 Grace--Blackwell Superchip (also sold as
the NVIDIA DGX Spark): a unified 20-core Arm64 CPU (10$\times$Cortex-X925
$+$ 10$\times$Cortex-A725) with an integrated Blackwell GPU (6{,}144
CUDA cores, 128~GB LPDDR5X unified memory). Reported wall-clock times
are end-to-end including data loading and plotting. The fixed random seed used throughout is
\texttt{seed=0} (PyTorch global generator); a multi-seed study of the
headline rows is reported in Supplementary~Sec.~S5. A reproducible
runtime environment is provided as an
\texttt{nvcr.io/nvidia/pytorch:26.01-py3} Docker recipe in the
repository \texttt{README}.

\section*{Competing interests}
The authors declare no competing interests.

\ack{The authors thank the University of Waterloo Department of Physics
\& Astronomy and Department of Chemistry for computational resources
and intellectual support during this work.}

\funding{No external funding was received for this work.}

\roles{C.G.T.F. conceived the method, implemented the \texttt{nsem} package,
designed and ran all experiments, analysed the results, prepared the
figures, and wrote the manuscript.
D.P. contributed to the scientific discussion and reviewed the manuscript.}

\data{All benchmark problems considered in this work use manufactured
analytic solutions whose definitions are reproduced in full in
Section~S2 of the Supplementary Information. The numerical outputs
underlying every figure and table --- training-loss curves,
per-iteration field snapshots, configuration files, and aggregated
ablation tables --- are available at
\url{https://github.com/cgtetsas/nsem} under the
directory \texttt{outputs/}, with run identifiers matching the
timestamps cited in figure captions and table footnotes. A permanent
Zenodo DOI will be minted at acceptance.}

\suppdata{Supplementary information is available for this paper at the
journal website. The supplementary file contains: theoretical
derivations of the NSEM framework (Section~S1), full non-dimensional
PNP system derivations and per-example boundary conditions (Section~S2),
additional benchmark results and solution panels (Sections~S3--S4),
and complete reproducibility commands with multi-seed statistics
(Section~S5).}

\bibliographystyle{iopart-num}
\bibliography{references}

\providecommand{\newblock}{}
\begin{thebibliography}{10}
\expandafter\ifx\csname url\endcsname\relax
  \def\url#1{{\tt #1}}\fi
\expandafter\ifx\csname urlprefix\endcsname\relax\def\urlprefix{URL }\fi
\providecommand{\eprint}[2][]{\url{#2}}

\bibitem{raissi2019pinn}
Raissi M, Perdikaris P and Karniadakis G~E 2019 {\em Journal of Computational
  Physics\/} {\bf 378} 686--707

\bibitem{karniadakis2021piml}
Karniadakis G~E, Kevrekidis I~G, Lu L, Perdikaris P, Wang S and Yang L 2021
  {\em Nature Reviews Physics\/} {\bf 3} 422--440

\bibitem{krishnapriyan2021failure}
Krishnapriyan A~S, Gholami A, Zhe S, Kirby R and Mahoney M~W 2021
  Characterizing possible failure modes in physics-informed neural networks
  {\em Advances in Neural Information Processing Systems (NeurIPS)\/} vol~34 pp
  26548--26560 (\textit{Preprint} \eprint{2109.01050})

\bibitem{tancik2020fourier}
Tancik M, Srinivasan P~P, Mildenhall B, Fridovich-Keil S, Raghavan N, Singhal
  U, Ramamoorthi R, Barron J~T and Ng R 2020 {Fourier} features let networks
  learn high-frequency functions in low-dimensional domains {\em Advances in
  Neural Information Processing Systems (NeurIPS)\/} vol~33 pp 7537--7547

\bibitem{rahaman2019spectralbias}
Rahaman N, Baratin A, Arpit D, Draxler F, Lin M, Hamprecht F~A, Bengio Y and
  Courville A 2019 On the spectral bias of neural networks {\em International
  Conference on Machine Learning (ICML)\/} (\textit{Preprint}
  \eprint{1806.08734})

\bibitem{mishra2023pinn}
Mishra S and Molinaro R 2023 {\em IMA Journal of Numerical Analysis\/} {\bf 43}
  1--43 (\textit{Preprint} \eprint{2006.16144})

\bibitem{mishra2022inverse}
Mishra S and Molinaro R 2022 {\em IMA Journal of Numerical Analysis\/} {\bf 42}
  981--1022 (\textit{Preprint} \eprint{2007.01138})

\bibitem{liu2024kan}
Liu Z, Wang Y, Vaidya S, Ruehle F, Halverson J, Solja\v{c}i\'{c} M, Hou T~Y and
  Tegmark M 2024 {\em Transactions on Machine Learning Research\/} ISSN
  2835-8856 (\textit{Preprint} \eprint{2404.19756})

\bibitem{patera1984spectral}
Patera A~T 1984 {\em Journal of Computational Physics\/} {\bf 54} 468--488

\bibitem{karniadakis2005spectral}
Karniadakis G~E and Sherwin S~J 2005 {\em Spectral/hp Element Methods for
  Computational Fluid Dynamics\/} 2nd ed (Oxford University Press)

\bibitem{komatitsch1999spectral}
Komatitsch D and Tromp J 1999 {\em Geophysical Journal International\/} {\bf
  139} 806--822

\bibitem{jagtap2020xpinn}
Jagtap A~D and Karniadakis G~E 2020 {\em Communications in Computational
  Physics\/} {\bf 28} 2002--2041 (\textit{Preprint} \eprint{2004.02518})

\bibitem{moseley2023fbpinn}
Moseley B, Markham A and Nissen-Meyer T 2023 {\em Advances in Computational
  Mathematics\/} {\bf 49} (\textit{Preprint} \eprint{2107.07871})

\bibitem{kharazmi2021hpvpinn}
Kharazmi E, Zhang Z and Karniadakis G~E 2021 {\em Computer Methods in Applied
  Mechanics and Engineering\/} {\bf 374} 113547 (\textit{Preprint}
  \eprint{2003.05385})

\bibitem{anandh2024fastvpinn}
Anandh T, Ghose D, Jain H and Ganesan S 2024 {\em SIAM Journal on Scientific
  Computing\/} {\bf 46} A3881--A3908 (\textit{Preprint} \eprint{2404.12063})

\bibitem{du2024nsm}
Du Y, Chalapathi N and Krishnapriyan A 2024 Neural spectral methods:
  Self-supervised learning in the spectral domain {\em International Conference
  on Learning Representations (ICLR)\/} (\textit{Preprint} \eprint{2312.05225})

\bibitem{yu2024sinn}
Yu T, Qi Y, Oseledets I and Chen S 2025 {\em Journal of Computational and
  Applied Mathematics\/} (\textit{Preprint} \eprint{2408.16414})

\bibitem{shukla2024neurosem}
Shukla K, Zou Z, Chan C~H, Pandey A, Wang Z and Karniadakis G~E 2024 {\em
  Computer Methods in Applied Mechanics and Engineering\/} {\bf 433} 117498
  (\textit{Preprint} \eprint{2407.21217})

\bibitem{anagnostopoulos2024pirbn}
Anagnostopoulos S~J, Toscano J~D, Stergiopulos N and Karniadakis G~E 2024 {\em
  Computer Methods in Applied Mechanics and Engineering\/} {\bf 421} 116805
  (\textit{Preprint} \eprint{2307.00379})

\bibitem{wang2024causal}
Wang S, Sankaran S and Perdikaris P 2024 {\em Computer Methods in Applied
  Mechanics and Engineering\/} {\bf 421} 116813 (\textit{Preprint}
  \eprint{2203.07404})

\bibitem{wang2024kinn}
Wang Y, Sun J, Bai J, Anitescu C, Eshaghi M~S, Zhuang X, Rabczuk T and Liu Y
  2024 {\em Computer Methods in Applied Mechanics and Engineering\/}
  (\textit{Preprint} \eprint{2406.11045})

\bibitem{zhang2024legendkinn}
Zhang Z, Xiong X, Zhang S, Wang W, Zhong Y, Yang C and Yang X 2025 {\em Expert
  Systems with Applications\/} (\textit{Preprint} \eprint{2406.08992})

\bibitem{ss2024chebkan}
SS S and R G 2024 {Chebyshev} polynomial-based {Kolmogorov--Arnold} networks:
  An efficient architecture for nonlinear function approximation arXiv preprint
  arXiv:2405.07200 (\textit{Preprint} \eprint{2405.07200})

\bibitem{sankaran2024chebpikan}
Guo C, Sun L, Li S, Yuan Z and Wang C 2024 Physics-informed
  {Kolmogorov--Arnold} network with {Chebyshev} polynomials for fluid mechanics
  arXiv preprint arXiv:2411.04516 (\textit{Preprint} \eprint{2411.04516})

\bibitem{berrut2004barycentric}
Berrut J~P and Trefethen L~N 2004 {\em SIAM Review\/} {\bf 46} 501--517

\bibitem{trefethen2000spectral}
Trefethen L~N 2000 {\em Spectral Methods in {MATLAB}\/} (Philadelphia: SIAM)

\bibitem{boyd2001spectral}
Boyd J~P 2001 {\em Chebyshev and {Fourier} Spectral Methods\/} 2nd ed (New
  York: Dover)

\bibitem{light1985dvr}
Light J~C, Hamilton I~P and Lill J~V 1985 {\em Journal of Chemical Physics\/}
  {\bf 82} 1400

\bibitem{colbert1992dvr}
Colbert D~T and Miller W~H 1992 {\em Journal of Chemical Physics\/} {\bf 96}
  1982--1991

\bibitem{wang2024adessential}
Wang Z, Hao S, Zhang Y and Zhang L 2024 {\em SIAM Journal on Numerical
  Analysis\/} {\bf 62} 1702--1720 (\textit{Preprint} \eprint{2405.14099})

\bibitem{kosloff1993kte}
Kosloff D and Tal-Ezer H 1993 {\em Journal of Computational Physics\/} {\bf
  104} 457--469

\bibitem{bernardi1994mortar}
Bernardi C, Maday Y and Patera A~T 1994 A new nonconforming approach to domain
  decomposition: The mortar element method {\em Nonlinear Partial Differential
  Equations and Their Applications, Coll\`ege de France Seminar, vol.~XI\/} ed
  Brezis H and Lions J~L (Pitman Research Notes in Mathematics) pp 13--51

\bibitem{lacour1997mortar}
Lacour C and Maday Y 1997 {\em BIT Numerical Mathematics\/} {\bf 37} 720--738

\bibitem{chen2024brdr}
Chen W, Howard A~A and Stinis P 2025 {\em Journal of Computational Physics\/}
  (\textit{Preprint} \eprint{2407.01613})
  \urlprefix\url{https://www.sciencedirect.com/science/article/pii/S0021999125005091}

\bibitem{liu1989lbfgs}
Liu D~C and Nocedal J 1989 {\em Mathematical Programming\/} {\bf 45} 503--528

\bibitem{byrd1995lbfgsb}
Byrd R~H, Lu P, Nocedal J and Zhu C 1995 {\em SIAM Journal on Scientific
  Computing\/} {\bf 16} 1190--1208

\bibitem{doumeche2023convergence}
Doum\`eche N, Biau G and Boyer C 2023 On the convergence of {PINNs} arXiv
  preprint arXiv:2305.01240 (\textit{Preprint} \eprint{2305.01240})

\bibitem{newman2004textbook}
Newman J and Thomas-Alyea K~E 2004 {\em Electrochemical Systems\/} 3rd ed
  (Wiley)

\bibitem{bard2001textbook}
Bard A~J and Faulkner L~R 2001 {\em Electrochemical Methods: Fundamentals and
  Applications\/} 2nd ed (Wiley)

\bibitem{bazant2004diffuse}
Bazant M~Z, Thornton K and Ajdari A 2004 {\em Physical Review E\/} {\bf 70}
  021506 (\textit{Preprint} \eprint{cond-mat/0401118})

\bibitem{huang2025epinn}
Huang X, Wang F, Zhang B and Liu H 2025 {\em Mathematics and Computers in
  Simulation\/} {\bf 237} 231--246 (\textit{Preprint} \eprint{2402.01768})

\bibitem{wang2022ntk}
Wang S, Yu X and Perdikaris P 2022 {\em Journal of Computational Physics\/}
  {\bf 449} 110768 (\textit{Preprint} \eprint{2007.14527})

\bibitem{macdonald1953ac}
Macdonald J~R 1953 {\em Physical Review\/} {\bf 92} 4--17

\bibitem{lasia2014eis}
Lasia A 2014 {\em Electrochemical Impedance Spectroscopy and Its
  Applications\/} 2nd ed (Springer)

\bibitem{orazem2017eis}
Orazem M~E and Tribollet B 2017 {\em Electrochemical Impedance Spectroscopy\/}
  2nd ed (Wiley (ECS Series))

\bibitem{doyle1993dfn}
Doyle M, Fuller T~F and Newman J 1993 {\em Journal of The Electrochemical
  Society\/} {\bf 140} 1526--1533

\end{thebibliography}


\providecommand{\newblock}{}
\begin{thebibliography}{10}
\expandafter\ifx\csname url\endcsname\relax
  \def\url#1{{\tt #1}}\fi
\expandafter\ifx\csname urlprefix\endcsname\relax\def\urlprefix{URL }\fi
\providecommand{\eprint}[2][]{\url{#2}}

\bibitem{karniadakis2005spectral}
Karniadakis G~E and Sherwin S~J 2005 {\em Spectral/hp Element Methods for
  Computational Fluid Dynamics\/} 2nd ed (Oxford University Press)

\bibitem{trefethen2000spectral}
Trefethen L~N 2000 {\em Spectral Methods in {MATLAB}\/} (Philadelphia: SIAM)

\bibitem{boyd2001spectral}
Boyd J~P 2001 {\em Chebyshev and {Fourier} Spectral Methods\/} 2nd ed (New
  York: Dover)

\bibitem{berrut2004barycentric}
Berrut J~P and Trefethen L~N 2004 {\em SIAM Review\/} {\bf 46} 501--517

\bibitem{kosloff1993kte}
Kosloff D and Tal-Ezer H 1993 {\em Journal of Computational Physics\/} {\bf
  104} 457--469

\bibitem{bernardi1994mortar}
Bernardi C, Maday Y and Patera A~T 1994 A new nonconforming approach to domain
  decomposition: The mortar element method {\em Nonlinear Partial Differential
  Equations and Their Applications, Coll\`ege de France Seminar, vol.~XI\/} ed
  Brezis H and Lions J~L (Pitman Research Notes in Mathematics) pp 13--51

\bibitem{lacour1997mortar}
Lacour C and Maday Y 1997 {\em BIT Numerical Mathematics\/} {\bf 37} 720--738

\bibitem{wang2022ntk}
Wang S, Yu X and Perdikaris P 2022 {\em Journal of Computational Physics\/}
  {\bf 449} 110768 (\textit{Preprint} \eprint{2007.14527})

\bibitem{liu2024kan}
Liu Z, Wang Y, Vaidya S, Ruehle F, Halverson J, Solja\v{c}i\'{c} M, Hou T~Y and
  Tegmark M 2024 {\em Transactions on Machine Learning Research\/} ISSN
  2835-8856 (\textit{Preprint} \eprint{2404.19756})

\bibitem{wang2024kinn}
Wang Y, Sun J, Bai J, Anitescu C, Eshaghi M~S, Zhuang X, Rabczuk T and Liu Y
  2024 {\em Computer Methods in Applied Mechanics and Engineering\/}
  (\textit{Preprint} \eprint{2406.11045})

\bibitem{zhang2024legendkinn}
Zhang Z, Xiong X, Zhang S, Wang W, Zhong Y, Yang C and Yang X 2025 {\em Expert
  Systems with Applications\/} (\textit{Preprint} \eprint{2406.08992})

\bibitem{ss2024chebkan}
SS S and R G 2024 {Chebyshev} polynomial-based {Kolmogorov--Arnold} networks:
  An efficient architecture for nonlinear function approximation arXiv preprint
  arXiv:2405.07200 (\textit{Preprint} \eprint{2405.07200})

\bibitem{sankaran2024chebpikan}
Guo C, Sun L, Li S, Yuan Z and Wang C 2024 Physics-informed
  {Kolmogorov--Arnold} network with {Chebyshev} polynomials for fluid mechanics
  arXiv preprint arXiv:2411.04516 (\textit{Preprint} \eprint{2411.04516})

\bibitem{light1985dvr}
Light J~C, Hamilton I~P and Lill J~V 1985 {\em Journal of Chemical Physics\/}
  {\bf 82} 1400

\bibitem{colbert1992dvr}
Colbert D~T and Miller W~H 1992 {\em Journal of Chemical Physics\/} {\bf 96}
  1982--1991

\bibitem{li2018losslandscape}
Li H, Xu Z, Taylor G, Studer C and Goldstein T 2018 Visualizing the loss
  landscape of neural nets {\em Advances in Neural Information Processing
  Systems (NeurIPS)\/} (\textit{Preprint} \eprint{1712.09913})

\bibitem{krishnapriyan2021failure}
Krishnapriyan A~S, Gholami A, Zhe S, Kirby R and Mahoney M~W 2021
  Characterizing possible failure modes in physics-informed neural networks
  {\em Advances in Neural Information Processing Systems (NeurIPS)\/} vol~34 pp
  26548--26560 (\textit{Preprint} \eprint{2109.01050})

\end{thebibliography}

\end{document}